\documentclass[aps,prb,twocolumn,amsmath]{revtex4-1} 
\pdfoutput=1
\usepackage{ulem}
\usepackage{dsfont}

\usepackage{color,graphicx}
\usepackage{bm}

\usepackage{amsmath}
\usepackage{amsfonts}
\usepackage{amssymb}

\usepackage{subfigure}
\usepackage[latin1]{inputenc}

\usepackage{balance}

\newcommand{\ket}[1]{\ensuremath{\left| #1 \right\rangle}}
\newcommand{\bra}[1]{\ensuremath{\left\langle #1 \right|}}
\newcommand{\sand}[2]{\left\langle #1| #2\right\rangle}

\newcommand{\etal}{{\it et al.}}

\renewcommand{\-}{\,-\,}

\newcommand{\br}{\mathbf{r}}

\newcommand{\bq}{\mathbf{q}}

\newcommand{\be}{\begin{equation}}
\newcommand{\ee}{\end{equation}}

\newcommand{\bea}{\begin{equation}\begin{aligned}}
\newcommand{\eea}{\end{aligned}\end{equation}}

\let\oldmarginpar\marginpar
\renewcommand\marginpar[1]{\-\oldmarginpar[\raggedleft\tiny #1]%
{\raggedright\tiny #1}}

\DeclareMathOperator{\Tr}{Tr}

\newcommand{\subfigimg}[3][,]{%
  \setbox1=\hbox{\includegraphics[#1]{#3}}
  \leavevmode\rlap{\usebox1}
  \rlap{\hspace*{-5pt}\raisebox{\dimexpr\ht1-2\baselineskip}{#2}}
  \phantom{\usebox1}
}

\begin{document}


\title{Exact solutions of fractional Chern insulators:  interacting particles in the Hofstadter model at finite size}
\author{Thomas Scaffidi}
\affiliation{Rudolf Peierls Centre for Theoretical Physics, Oxford OX1 3NP, United Kingdom}

\author{Steven H. Simon}
\affiliation{Rudolf Peierls Centre for Theoretical Physics, Oxford OX1 3NP, United Kingdom}

\date{\today}

\begin{abstract}

We show that all the bands of the Hofstadter model on the torus have an exactly flat dispersion and Berry curvature when a special system size is chosen.
This result holds for any hopping and Chern number.
%
%
%
%
Our analysis therefore provides a simple rule for choosing a particularly advantageous system size when designing a Hofstadter system whose size is controllable, like a qubit lattice or an optical cavity array. 
The density operators projected onto the flat bands obey exactly the Girvin-MacDonald-Platzman algebra, like for Landau levels in the continuum in the case of $C=1$, or obey its straightforward generalization in the case of $C>1$.
%
%
This allows a mapping between density-density interaction Hamiltonians for particles in the Hofstatder model and in a continuum Landau level. 
By using the well-known pseudopotential construction in the latter case, we obtain fractional Chern insulator phases, the lattice counterpart of fractional quantum Hall phases, that are exact zero-energy ground states of the Hofstadter model with certain interactions.
%
Finally, the addition of a harmonic trapping potential is shown to lead to an appealingly symmetric description in which a new Hofstadter model appears in momentum space.


\end{abstract}


\maketitle

\section{Introduction}
Fractional quantum Hall (FQH) phases are the archetype of the topological phase of matter. 
They are interesting both from a fundamental perspective and as a resource for topological quantum computation\cite{RevModPhys.80.1083}. 
While FQH phases were originally defined in terms of electrons in a Landau level, it was shown that such phases can also exist in the presence of a periodic potential \cite{PhysRevB.48.8890} and for atoms in optical lattices \cite{PhysRevLett.94.086803,PhysRevA.76.023613,PhysRevLett.96.180407,PhysRevA.78.013609,PhysRevLett.103.105303,PhysRevB.86.165314}. 
More recently, a flurry of numerical studies have shown that gapped topological phases resembling the FQH phases can also be obtained in a variety of lattice models with topological flat bands (TFBs).
These phases were baptized fractional Chern insulators (FCIs) (see Refs.~ \cite{PhysRevB.85.075116, Parameswaran2013816, doi:10.1142/S021797921330017X} and references therein).

The Hofstadter model \cite{azbel1964energy, 0370-1298-68-10-305, PhysRevB.14.2239} occupies a special place in the large set of known lattice models exhibiting TFBs. 
It is one of the most studied of these models, it is very simple to define --- a square lattice with nearest-neighbor hopping and uniform flux density --- but it leads to very rich physics, as exemplified by the Hofstadter butterfly \cite{PhysRevB.14.2239} giving the intricate dependence of the spectrum on the flux density.
There is also a large number of proposals for its experimental realization with cold atoms in optical lattices \cite{PhysRevLett.108.225303, PhysRevLett.108.225304, PhysRevLett.111.225301, PhysRevLett.111.185301, PhysRevLett.111.185302} and two of them were recently carried out successfully \cite{PhysRevLett.111.185301, PhysRevLett.111.185302}.
While these two experiments have focused on single-particle physics, interesting many-body physics in such systems, including FQH-like phases \cite{PhysRevLett.94.086803,PhysRevA.76.023613,PhysRevLett.96.180407,PhysRevA.78.013609,PhysRevLett.103.105303, PhysRevB.86.165314} and other interesting possibilities \cite{PhysRevLett.104.145301, PhysRevLett.104.255303,PhysRevLett.108.045306, PhysRevLett.109.205303,2014arXiv1402.6704W}, might become reality in the near future.


%


Several approaches were used to study FCIs (see Refs. \cite{PhysRevB.85.075116, Parameswaran2013816, doi:10.1142/S021797921330017X} and references therein). 
Since most of our understanding of the fractional quantum Hall effect (FQHE) is based on its description in terms of electrons in a Landau level in a continuum system, it is highly desirable to relate the lattice problem to the continuum one.
Using a Wannier construction \cite{PhysRevLett.107.126803}, the FCIs were shown to be topologically equivalent to their continuum counterpart \cite{PhysRevLett.109.246805, PhysRevB.87.035306} and high overlaps between the two states were obtained \cite{PhysRevB.86.085129}.


The difference between Landau levels (LLs) and TFBs in lattice models lies in three different properties: the dispersion, the Berry curvature and the form factor.
The first is the simplest one: while Landau levels are energetically flat, TFBs will generally have some dispersion, although always smaller than the gap to the next band. 
Second, while the Berry curvature is flat over the Brillouin zone (BZ) for Landau levels, it is generally not the case for lattice models. The Berry curvature is related to the phase of the overlap between the periodic part of the different Bloch wave functions inside the TFB.
Third, the form factor, which modulates the amplitude of the projected interaction with a momentum space dependent factor, is related to the norm of this overlap.
The norm of this overlap can be interpreted as a ``distance'' in the manifold of single-particle states obtained inside one band after projection\cite{PhysRevLett.107.116801}.
Whereas LL form factors are given by a Laguerre polynomial multiplied by a Gaussian, they are non generic and have to be computed numerically for most lattice models.
In the case of the LLL, Haldane showed that the form factor is related to a ``Galilean'' metric\cite{PhysRevLett.107.116801}, which, when assumed rotationally invariant, conceals a fundamental geometric degree of freedom of FQH fluids.

In summary, lattice models will generally differ from Landau levels in these three ways at the same time: the bands will have a non flat dispersion and Berry curvature, and the form factor will differ from the Gaussian function found in the LLL case.
Besides, while Landau levels all have $C=1$, lattice TFBs can have $|C|>1$.
The addition of interaction to these bands can lead to new intriguing color-entangled FQH-like phases at different filling fractions than for LLs\cite{PhysRevLett.103.105303,PhysRevLett.108.256809,PhysRevB.87.205137,PhysRevLett.110.106802, PhysRevB.89.155113,PhysRevLett.111.186804}.

In order to study the minimal ingredients required to obtain FQH-like phases on lattices and to make their resemblance to their continuum counterparts explicit, it would be beneficial to construct models for which one can vary at will the ``closeness'' to Landau levels in one way while keeping the two others fixed. In this work, we provide such a model, by showing that, at the right system size and flux density, the Hofstatder model on the torus provides bands with exactly flat dispersion and Berry curvature. The only difference with LLs is therefore the form factor of each band, i.e., the ``distances'' between single-particle states inside each band.


Furthermore, for our system size, the density operators projected to any band of the Hofstadter model obey exactly the same algebra as in the continuum --- the Girvin-MacDonald-Platzman (GMP) algebra\cite{PhysRevB.33.2481} (or a simple generalization thereof in the case of $C>1$).
This algebra, obeyed only approximately for generic lattice models with TFBs, was successfully used to study FCIs\cite{ PhysRevB.85.241308,2012arXiv1208.2055R, PhysRevB.86.195146, 2011arXiv1108.5501M}.
The GMP algebra is central to the description of FQH phases in terms of guiding-center degrees of freedom only\cite{PhysRevLett.107.116801, PhysRevB.87.115103, PhysRevB.88.165303}, as opposed to the usual description of the FQHE based on first-quantized real-space many-body wave functions with the right vanishing properties like the Laughlin state.
These wave functions rely on the existence of analytical properties of the continuum LL single-particle states on the complex plane, which is a result of the interplay between both the guiding-center and the cyclotron dynamics.
As explained recently\cite{PhysRevLett.107.116801, PhysRevB.87.115103, PhysRevB.88.165303}, this situation is somewhat unsatisfactory since (1) it is very specific to the LLs and is not directly applicable to TFBs other than LLs and (2) the topological order of FQH phases is entirely contained in the guiding center degrees of freedom and a ``guiding-center only'' description of the FQHE is therefore highly desirable.
By providing a system different from LLs with projected densities obeying exactly the GMP algebra, we believe our work might be useful to generate progress in this direction.

%

%
%
%

Let us now summarize the main idea of the paper.
Our analysis starts from a very simple observation about the Hofstadter model.
In the Landau gauge, the momentum in the $y$ direction is a good quantum number and, for each $p_y$, the dynamics is governed by a one-dimensional chain of sites in the $x$ direction with nearest-neighbor hopping and a cosine potential (i.e., the Harper equation\cite{0370-1298-68-10-305}). 
More precisely, if we index the chain by $x \in \mathbb{Z}$, this potential is given by $\cos(p_y + 2 \pi x n_{\phi})$ where $n_{\phi}=P/Q$ is the flux density and $P$ and $Q$ are coprime.
The observation is the following: if we restrict $p_y$ to take values given by $2 \pi K_y/Q$ where $K_y \in \mathbb{Z}$, then increasing $K_y$ by a certain integer is equivalent to increasing $x$ by the same number.
This means that all the different Harper equations, each corresponding to a different value of $K_y$, would be equivalent up to a translation in $x$, and would therefore have the same spectrum, leading to exactly flat bands.
The way to restrict $p_y$ to these values is to use periodic boundary conditions (PBCs) in both directions and to take a finite-size system of $N_y=Q$ unit cells in the $y$ direction, or in other words a square system of $Q$ by $Q$ plaquettes.
All the other results then follow from this finite-size ``additional symmetry'', $N_y=Q$.

We should mention that a different finite-size effect leading to flat bands in the Hofstadter model was reported previously\cite{0295-5075-91-5-57007}. This effect occurs in a different setting: it appears on a thin torus with periodic boundary conditions enforced in the diagonal directions.


While our analysis only works for strictly finite-size systems, one can reach arbitrarily large systems at the expense of choosing a large $Q$ (which does not necessarily mean vanishing flux density since one can also choose a large $P$).
At any rate, finite-size systems are interesting in their own right, since there were several proposals for the realization of the Hofstadter model with qubit lattices\cite{PhysRevA.87.062336} and optical cavity arrays\cite{PhysRevA.84.043804}.
The stabilization of FQH phases in these systems was also discussed\cite{PhysRevX.4.031039}.
Our work is particularly suited for these systems since they would be inherently finite-sized (one can imagine a lattice of tens of qubits) and it would be possible to implement periodic boundary conditions through appropriate wiring.
Our analysis therefore provides a simple rule for choosing a particularly advantageous system size when designing such systems.



This work is organized as follows. 
In Sec. II A, we show the exact flatness of the dispersion and the Berry curvature for the aforementionned system size.
In Sec. II B, we take advantage of this flatness to write explicitly the projection of a generic density-density interaction to a given flat band of the Hofstadter model.
For simplicity, we initially focus on the case of $C=1$ and of a contact interaction projected to the lowest band.
We show an explicit link between the projected interaction and pseudopotentials projected to a continuum LL.
For the sake of definiteness, we will discuss mostly the bosonic Laughlin state as a ground state of a contact interaction projected to the lowest Landau level (LLL), but our analysis could easily be applied to any FQH state for which there is a parent Hamiltonian.
In Sec. III, we discuss the relation of our analysis to three previous works: (a) the Wannier construction for FCIs developed by Qi\cite{PhysRevLett.107.126803}, (b) the Kapit-Mueller model\cite{PhysRevLett.105.215303}, a model that exhibits an exactly flat band by using exponentially decaying hoppings, and (c) the impossibility of exactly flat Chern bands with strictly short-range hoppings shown by Chen \etal\cite{1751-8121-47-15-152001}.
In Sec. IV, we generalize our result to: (a) the case of higher Chern numbers, (b) the case of higher bands, (c) the case of pseudopotentials corresponding to other than point contact interactions, and (d) the case of a trapping harmonic potential.
We conclude in Sec. V with a brief summary of our results.
Appendix A is referred to in Sec. III B and gives the derivation of the continuum LLL wave functions sampled on a lattice.
Appendix B is referred to in Sec. III B and gives the calculation of the overlaps of these wave functions.


%
%

%
%
%


\section{Main result}
\subsection{Single-particle}
We study the Hofstatder model\cite{azbel1964energy, 0370-1298-68-10-305, PhysRevB.14.2239}: particles hopping on a square lattice $(x,y) \in (\mathbb{Z},\mathbb{Z})$ with a flux density of $-n_{\phi}=-P/Q$ per plaquette ($P$ and $Q$ are coprime). The Hamiltonian is given by
\begin{equation}
H_0 = -t_x (\hat{T}_x + \hat{T}_{-x})  -t_y (\hat{T}_y + \hat{T}_{-y}),
\end{equation}
where, in the Landau gauge,
\begin{equation}
\begin{aligned}
\hat{T}_x & = \sum_{x,y} \ket{x+1,y}\bra{x,y}, \\
\hat{T}_y & = \sum_{x,y} e^{-i 2 \pi \frac{P}{Q} x} \ket{x,y+1}\bra{x,y}\text{.}
\label{TOp}
\end{aligned}
\end{equation}

We now consider the following magnetic translation operators\cite{PhysRev.134.A1602, PhysRev.134.A1607, 0022-3719-18-22-004, RevModPhys.82.1959, 0253-6102-50-2-48, PhysRevLett.104.145301, PhysRevLett.104.255303}:
\begin{equation}
\begin{aligned}
T_x &= \sum_{x,y}  e^{-i 2 \pi \frac{P}{Q} y} \ket{x+1,y}\bra{x,y}, \\
T_y &= \sum_{x,y}  \ket{x,y+1}\bra{x,y}\text{.}
\end{aligned}
\end{equation}
It is easy to see that 
\begin{equation}
\begin{aligned}
T_x T_y & = T_y T_x e^{-i 2 \pi \frac{P}{Q}}, \\
\hat{T}_x \hat{T}_y & =\hat{T}_y \hat{T}_x e^{i 2 \pi \frac{P}{Q}}, \\
[T_{\mu}, \hat{T}_{\nu}] &=0 \ \forall \mu,\nu, \\
[T_x^Q, T_y] &=0\text{.}
\end{aligned}
\end{equation}
The last equation defines a unit cell of dimension $(Q,1)$ for which we will define Bloch momenta $p_x, p_y$.

The operators $T_{x,y}$ commute with $H_0$ and live in guiding center space: they couple states with different Bloch momenta, but within the same band.
In contrast, the operators $\hat{T}_{x,y}$ live in cyclotron space: they couple states from different bands but with the same momentum.
They will be projected out when treating the interaction between electrons inside one band.



Let $\ket{p_x,p_y,E}$ be an eigenstate of $T_x^{Q}$, $T_y$ and $H_0$:
\begin{equation}
\begin{aligned}
T_x^Q \ket{p_x,p_y,E} &= e^{- i p_x}  \ket{p_x,p_y,E}, \\
T_y \ket{p_x,p_y,E} &= e^{- i p_y} \ket{p_x,p_y,E}, \\
H_0 \ket{p_x,p_y,E} &= E \ket{p_x,p_y,E}\text{.}
\end{aligned}
\end{equation}
Using Bloch theorem, this state can be written as
\begin{equation}
\ket{p_x,p_y,E} =   \sum_{x,y} e^{i p_y y} e^{i p_x \frac{x}{Q}} u_{p_x,p_y,E}(x) \ket{x,y},
\end{equation}
where $ u_{p_y,n}(x) =  u_{p_y,n}(x+Q)$. 
Now, we apply $T_x$ on this state:
\begin{equation}
\begin{aligned}
&T_x \ket{p_x,p_y,E} \\
&=   \sum_{x,y} e^{-i 2 \pi n_{\phi} y} e^{i p_y y} e^{i p_x \frac{x-1}{Q}} u_{p_x,p_y,E}(x-1) \ket{x,y} \\
 &= e^{- i \frac{p_x}{Q}} \ket{p_x,p_y-2\pi n_{\phi}, E},
 \end{aligned}
\end{equation}
where, when going from the second to the third line, we used the fact that $T_x$ commutes with $H_0$.

By successive product of $T_x$ on a state $\ket{p_x,p_{y0},E}$, it is possible to generate a set of $Q$ degenerate states indexed by $s$:
\begin{equation}
\ket{p_x,p_{y0} -2 \pi n_{\phi} s,E}\text{.}
\end{equation}

We impose periodic boundary conditions (PBCs): $T_y^{N_y} = 1$ and $T_x^{Q N_x} = 1$, where $N_{x(y)}$ is the number of unit cells in the direction $x$($y$). This amounts to the restriction of $p_y = \frac{2 \pi}{N_y} k_y$ and  $p_x = \frac{2 \pi}{N_x} k_x$ with $k_x,k_y \in \mathbb{Z}$.


Crucially, if $N_x=1$ and $Q$ is a multiple of $N_y$, all the $N_y$ different values of $p_y$ that are allowed by the PBCs are generated by successive product of $T_x$ on $\ket{0,0,E}$:
\begin{equation}
\ket{P s,E} \equiv \ket{0,P s \frac{2 \pi}{N_y},E} = T_x^{-s} \ket{0,0,E}\text{.}
\end{equation}
Since $P$ and $Q$ are coprime, so are $P$ and $N_y$. This means that $Ps \mod N_y$ takes all its possible values $0 \dots N_y-1$.

More precisely,
\begin{equation}
\ket{K_y,E}=  T_x^{-\alpha C K_y} \ket{0,0,E}
\end{equation}
where $K_y=0 \dots N_y-1$, $C$ is a solution of the Thouless-Kohmoto-Nightingale-den Nijs\cite{PhysRevLett.49.405} (TKNN) Diophantine equation $P C \mod Q = 1$, and $\alpha$ is given by
\begin{equation}
\alpha \equiv \frac{Q}{N_y} \\ \in \mathbb{Z}\text{.}
\end{equation} 
For each value of $E$, we thus obtain $N_y$ states that are exactly degenerate and obey the PBCs.

\begin{figure}
  \centering
  \begin{tabular}{@{}p{0.45\linewidth}@{\quad}p{0.45\linewidth}@{}}
    \subfigimg[width=\linewidth]{a)}{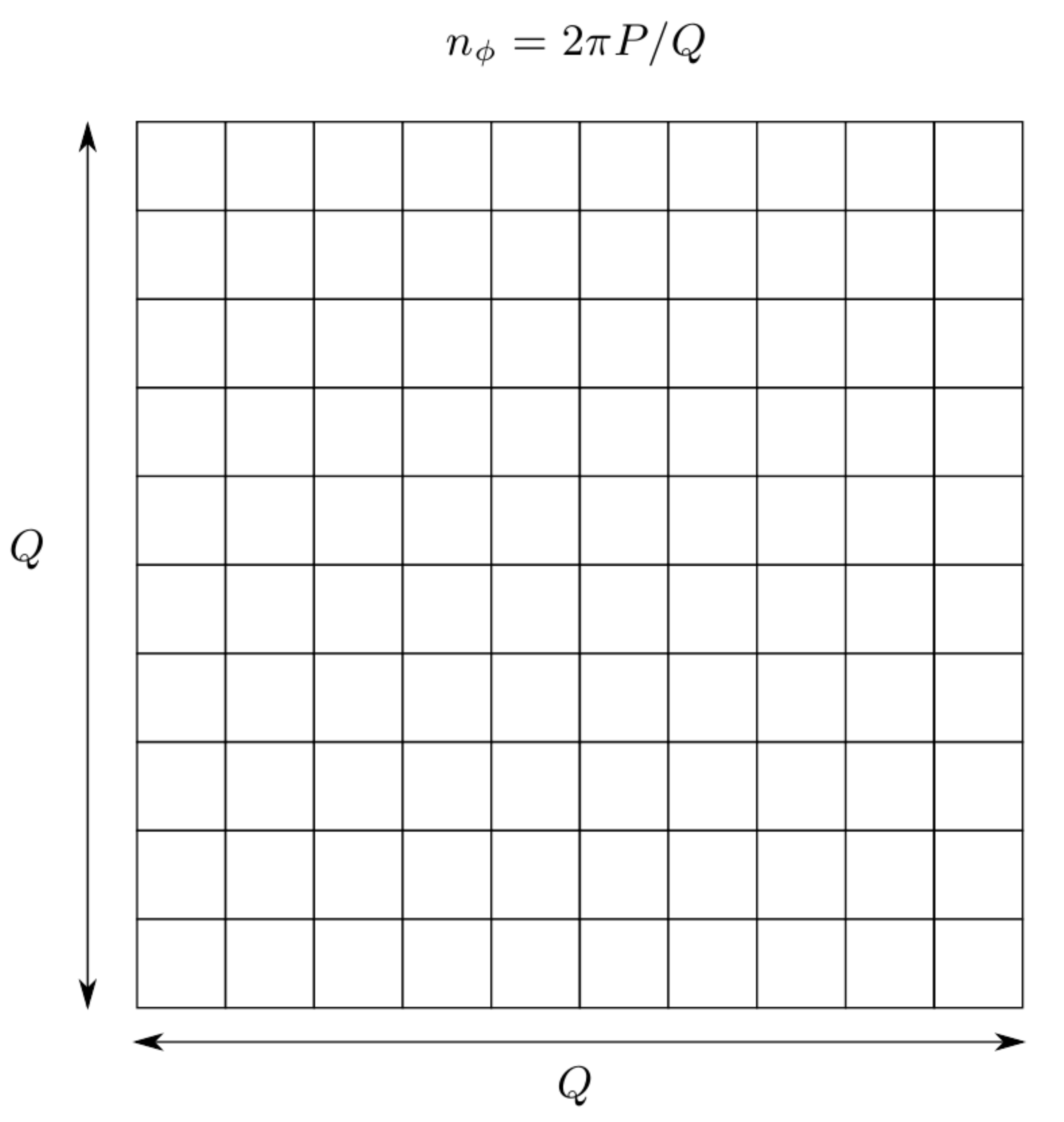} & 
    \subfigimg[width=\linewidth]{b)}{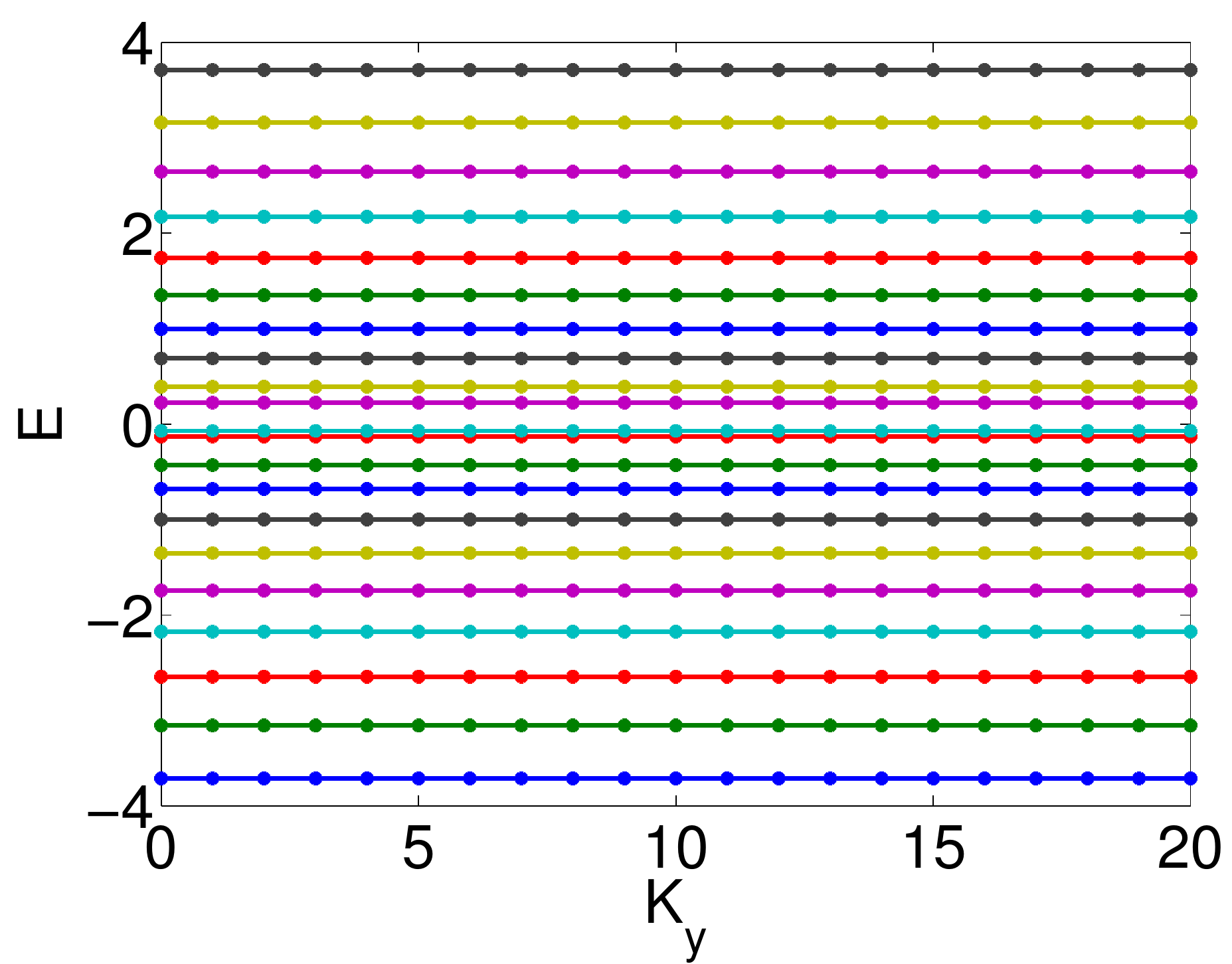} \\
  \end{tabular}
\caption{\label{OneBodySpec}
(a) Illustration of the system discussed in this work: a square lattice of $Q$ by $Q$ plaquettes with a uniform flux density of $n_{\phi}=2\pi P/Q$ and periodic boundary conditions. (b) Single-particle spectrum for $Q=21$. All the bands are exactly flat.
}
\end{figure}

The spectrum for $E$ is given by the $Q$ solutions $E_n$ of the Harper equation\cite{0370-1298-68-10-305}:
\begin{equation}
\begin{aligned}
E_n u_{K_y,n}(x) &=  2 \cos\left[2 \pi \left(\frac{K_y}{N_y} + x \frac{P}{Q}\right)\right] u_{K_y,n}(x) \\
&+u_{K_y,n}(x+1) + u_{K_y,n}(x-1). 
\end{aligned}
\end{equation}
We therefore obtain exactly flat bands of $N_y$ degenerate states each, situated at energy $E_n$, $n=1 \dots Q$. 
We note these states 
\begin{equation}
\begin{aligned}
\ket{K_y,n} &=  \sum_{x,y} e^{i K_y \frac{2 \pi}{N_y} y} u_{0,K_y \frac{2 \pi}{N_y},E_n}(x) \ket{x,y} \\
  & \equiv  \sum_{x,y} e^{i K_y \frac{2 \pi}{N_y} y} u_{K_y,n}(x) \ket{x,y}.
 \end{aligned}
\end{equation}
These exactly flat bands can be seen in Fig.~\ref{OneBodySpec} for the case of $\alpha=1$, along with an illustration of the system.
%
%

We now want to project the interaction between electrons onto one of the exactly flat bands.
We focus on density-density interaction Hamiltonians:
\begin{equation}
H_{\text{Lat}} = \sum_{\bq \in \text{BZ}} V({\bf q}) \tilde{\rho}_{\bf q} \tilde{\rho}_{-\bf q}, 
\end{equation}
where $ \tilde{\rho}_{\bf q}$ is the density operator:
\begin{equation}
 \tilde{\rho}_{\bf q} = \sum_{x,y}  e^{i 2 \pi (q_x \frac{x}{Q} + q_y \frac{y}{N_y})} \ket{x,y}\bra{x,y} 
\end{equation}
and where ${\bf q}$ is summed over a BZ given by $(q_x,q_y)$ with $q_x = 0 \dots Q-1$ and $q_y = 0 \dots N_y-1$, since our system size corresponds to a system of $Q = \alpha N_y$ by $N_y$ plaquettes.
We should emphasize that $\bq$ belongs to an ``enlarged'' Brillouin zone compared to the magnetic Brillouin zone of the single-particle states, $\ket{K_y}$, which, for our system size, is given by $(K_x, K_y)$ with $K_x=0$ and $K_y=0,\dots,N_y-1$.




Crucially, if we define $\tilde{\rho}_x \equiv  \tilde{\rho}_{(1,0)}$ and $\tilde{\rho}_y \equiv  \tilde{\rho}_{(0,1)}$, we have
\begin{equation}
\begin{aligned}
\tilde{\rho}_x &=  T_y^C \hat{T}_y^{-C}, \\
\tilde{\rho}_y &=  T_x^{-\alpha C} \hat{T}_x^{\alpha C}.
 \end{aligned}
\end{equation}
This means that, for our system size, the density operator is a product of an operator in guiding center space and one in cyclotron space, just like for Landau levels in the continuum in which case the density operators are given by $ \tilde{\rho}_{\bf q} = e^{i {\bf q} \cdot \hat{ \bf r}} =  e^{i {\bf q} \cdot {\bf \hat{R}} } e^{i {\bf q} \cdot \hat{\boldsymbol\eta}} $ with $ \hat{ \bf r} = {\bf \hat{R}} + \hat{\boldsymbol\eta}$ the position operator, ${\bf \hat{R}}$ the guiding center position operator and $\hat{\boldsymbol\eta}$ the cyclotron position operator.

The projection operator onto band $n$ lies in cyclotron space (it is made of the $\hat{T}$ operators) and therefore commutes with the $T$ operators. 
We obtain
\begin{equation}
 \rho_{x,n} \equiv P_n \tilde{\rho}_{x} P_n =  T_y^C \ P_n \hat{T}_y^{-C}P_n \equiv  T_y^C \ d_n(1,0),
\end{equation}
where $d_n(1,0)$ is simply a scalar depending on the periodic part of the Bloch wavefunction $u_{K_y,n}(x)$:
\begin{equation}
d_n(1,0) = \sum_x u_{0,n}^*(x) u_{0,n}(x) e^{i 2 \pi \frac{x}{Q}}\text{.}
\end{equation}
Similarly, we obtain
\begin{equation}
 \rho_{y,n} \equiv P_n \tilde{\rho}_{y} P_n =  T_x^{- \alpha C} \ P_n \hat{T}_x^{\alpha C} P_n \equiv  T_x^{- \alpha C} \ d_n(0,1),
\end{equation}
where $d_n(0,1)$ is given by 
\begin{equation}
d_n(0,1) = \sum_x u_{0,n}^*(x) u_{0,n}(x+\alpha C)\text{.}
\label{distanceHaldane}
\end{equation}
From Eq.~(\ref{distanceHaldane}), we see that $d_n(0,1)$ is given by the distance in the sense of Ref.~\cite{PhysRevLett.107.116801} between two Bloch wave functions differing by one unit in the $K_y$ direction.
%

We can now compute the finite system version of the Chern number of band $n$, defined by\cite{PhysRevLett.110.106802}
\begin{equation}
\begin{aligned}
\boldsymbol{C}_n &= \frac1{2\pi} \Tr \operatorname{Im} \ln[  \rho_{x,n}  \rho_{y,n}  \rho_{x,n}^{-1} \rho_{y,n}^{-1}] \\
&= \frac1{2\pi} \Tr \operatorname{Im} \ln[  T_y^C T_x^{- \alpha C}  T_y^{-C} T_x^{\alpha C}] \\
& = \frac1{2\pi} \sum_{K_y} \mathcal{B}(K_y),
\label{RK}
\end{aligned}
\end{equation}
where the finite system version of the Berry curvature is given by the phase of Wilson loops in momentum space:
\begin{equation}
\mathcal{B}(K_y) = \operatorname{Im} \ln[  \bra{K_y} T_y^C T_x^{- \alpha C}  T_y^{-C} T_x^{\alpha C} \ket{K_y}]\text{.} \end{equation}

Using $T_x^{- \alpha C}  T_y^{-C} = T_y^{-C} T_x^{- \alpha C} e^{i 2 \pi \frac{P}{Q} C^2 \alpha}$, we find that the Berry curvature is uniform in $K_y$ space, $\mathcal{B}(K_y) = \alpha \frac{P}{Q} C^2 2 \pi$, and that the Chern number is given by 
\begin{equation}
\begin{aligned}
\boldsymbol{C}_n &= N_y \alpha \frac{P}{Q} C^2  = P C^2 \\
&= C \mod Q,
\end{aligned}
\end{equation}
 where we used the fact that $PC \mod Q = 1$. 
We find back the well-known result that all the bands of the Hofstadter model have the same Chern number modulo $Q$.
For our system size, the discrete set of $K_y$ points does not allow a continuous pumping of $p_y$ to resolve the modulo $Q$ ambiguity, and the Chern number is thus only defined modulo $Q$. 

The equality between the first two lines of Eq.~(\ref{RK}) shows that, for our system size, there is an explicit link between the Wilson loops in real space and the ones in momentum space: a Wilson loop of one ``plaquette'' in $K_y$ space corresponds to a Wilson loop of $C^2$ plaquettes in real space.

We now turn to the projection of the density operator for general ${\bf q}=(q_x,q_y)$:
\begin{equation}
\begin{aligned}
 \rho_{{\bf q}} \equiv P_n \tilde{\rho}_{{\bf q}} P_n &=  T_y^{C q_x} \ T_x^{-\alpha C q_y} \ P_n \ \hat{T}_y^{-C q_x} \ \hat{T}_x^{\alpha C q_y} \ P_n \\
& \equiv   \bar\rho_{{\bf q}} \  d_n(q_x,q_y)
\end{aligned}
\end{equation}
with
\begin{equation}
\begin{aligned}
 \bar\rho_{{\bf q}} & \equiv T_y^{C q_x} \ T_x^{-\alpha C q_y} \ e^{i \frac12 \frac{2 \pi}{N_y} C q_x q_y}, \\
d_n(q_x,q_y) \mathds{1} &\equiv P_n \ \hat{T}_y^{-C q_x} \ \hat{T}_x^{\alpha C q_y} \ P_n e^{-i \frac12 \frac{2 \pi}{N_y} C q_x q_y},
\end{aligned}
\end{equation}
where the complex phase is added only as a matter of convention and is chosen so that $\bar{\rho}_{{\bf q}}$ obeys the following algebra (we will refer to it as the $C$-GMP algebra since it is the straightforward generalisation of the GMP algebra to any Chern number $C$):
\begin{equation}
\begin{aligned}
 \left[\bar\rho_{{\bf q}_1}, \bar\rho_{{\bf q}_2} \right]  &= \bar\rho_{{\bf q}_1+{\bf q}_2} 2 i \sin\left(-\frac12  \frac{2 \pi}{N_y} C (q_{1x} q_{2y} - q_{1y} q_{2x}) \right) \\
& \equiv  \bar\rho_{{\bf q}_1+{\bf q}_2} 2 i \sin\left(-\frac12  \frac{2 \pi}{N_y} C ({\bf q}_1 \times {\bf q}_2) \right)\text{.}
\end{aligned}
\end{equation}

For the sake of completeness, we give the action of these operators in the $\ket{K_y}$ basis:
\begin{equation}
\bar\rho_{{\bf q}} \ket{K_y} =  e^{-i \frac12 \frac{2 \pi}{N_y} C q_x q_y} e^{-i \frac{2 \pi}{N_y} C K_y q_x} \ket{K_y+q_y} \text{.}
\label{rhoGMPdef}
\end{equation}
The distance is given by
\begin{equation}
\begin{aligned}
d_n(q_x,q_y) &=  e^{-i \frac12 \frac{2 \pi}{N_y} C q_x q_y} \sum_x  e^{i 2 \pi \frac{x}{Q} q_x} \times\\
&u_{0,n}^*(x) u_{0,n}(x+\alpha C q_y)  
\label{distance}
\end{aligned}
\end{equation}
and we have $d_n(-q_x,-q_y) =  d_n^*(q_x,q_y)$.


\subsection{Projected Interaction}
For the reminder of the paper, we will restrict ourselves to the case $\alpha=1$ (i.e., $Q=N_y$) in order to simplify notations, but one could easily generalize all the results for $\alpha \neq 1$. For the same reason, we will consider only the lowest band for now and drop the $n$ index, but we will generalize to higher bands in Sec. IV.

%


We can now write the Hamiltonian projected to the lowest band as
\begin{equation}
\bar{H}_{\text{Lat}} \equiv P_0 H_{\text{Lat}} P_0 = \sum_{\bq \in \text{BZ}}  F_{\text{Lat}}({\bf q})  \ {\bar\rho}_{\bf q} {\bar\rho}_{-\bf q},
\label{PHP1}
\end{equation}
where
\begin{equation}
F_{\text{Lat}}({\bf q}) = V({\bf q}) |d_0({\bf q})|^2
\end{equation}
and where the BZ is given by $q_{x}=0,\dots,N_y-1$ and $q_{y}=0,\dots,N_y-1$.
This expression for the projected Hamiltonian neatly separates the universal, guiding-center part --- the GMP operators ${\bar\rho}_{\bf q}$ which are exactly the same as in the continuum --- from the non universal, cyclotron part --- the form factor $|d_0({\bf q})|^2$ that depends on the specific band and hoppings and is related to the Galilean metric of the band\cite{PhysRevLett.107.116801}.
Our results are true for any hopping Hamiltonian (and therefore for any hopping range), as long as it is written in terms of $\hat{T}_{x,y}$ and their products (which simply amounts to requiring that the phase of each hopping is consistent with the flux density per plaquette $P/Q$).
By varying the hoppings, one can therefore engineer the form factor $|d_0({\bf q})|^2$ of a flat band without changing any other properties, namely the dispersion and the Berry curvature, which remain exactly the same as for a Landau level.

Our results are also applicable to periodically driven hopping Hamiltonians written in terms of $\hat{T}_{x,y}$ and their products, like the system studied in Ref.~\cite{PhysRevLett.112.026805}. 
These systems have also been shown to host FQH-like phases, deemed Floquet fractional Chern insulators\cite{PhysRevLett.112.156801}.

We will now show an explicit link between the Hamiltonian in Eq.~(\ref{PHP1}) and its equivalent for electrons in the continuum torus.
The interaction Hamiltonian of the $m$th pseudopotential (which is the torus analog of the interaction on the plane that projects to the space of relative angular momentum $m$) projected to the LLL is given by\cite{PhysRevLett.107.116801}
\begin{equation}
\bar{H}_c \equiv P_{\text{LLL}} H_{\text{LLL}} P_{\text{LLL}} = \sum_{\bq \in \text{BZ}}  F_{\text{LLL}}^{(m)} ({\bf q})  \ \bar\rho_{{\bf q}} \bar\rho_{-{\bf q}},
\label{PHPc1}
\end{equation}
where 
\be
F^{(m)}_{\text{LLL}}({\bf q}) = \sum_{{\bf H}}  V^{(m)}(({\bf q}+{\bf H})^2 l^2)  e^{- \frac12 ({\bf q}+{\bf H})^2 l^2}
\ee
with $V^{(m)}(({\bf q}+{\bf H})^2 l^2) = L^{m}(({\bf q}+{\bf H})^2 l^2)$ the $m$th Laguerre polynomial.

In Eq.~(\ref{PHPc1}), the sum is over the same $N_y \times N_y$ BZ as in Eq.~(\ref{PHP1}) and the operators $\bar\rho_{{\bf q}}$ have exactly the same matrix elements as the ones defined in Eq.~(\ref{rhoGMPdef}) if one identifies the continuum LLL single-particle states $\ket{\mathcal{K}_y}$ (defined for  $\mathcal{K}_y = 0,\dots,N_{y}-1$) with their lattice counterpart $\ket{K_y}$.
The states $\ket{\mathcal{K}_y}$ are given by
\bea
\sand{\tilde{x}, \tilde{y}}{\mathcal{K}_y} &= \sum_{n \in \mathbb{Z}}  \exp\left[ i \frac{2 \pi}{L_y} (\mathcal{K}_y+ n N_{y}) \tilde{y}  \right] \\
& \times \exp\left[ -\frac{\left(\tilde{x} + l^2 (\mathcal{K}_y+  n N_{y}) \frac{2 \pi}{L_y}\right)^2  }{2 l^2} \right], 
\label{OneBodyStatesKy}
\eea
where $\tilde{x},\tilde{y}$ are the spatial coordinates on the torus and $l$ is the magnetic length.
The torus is pierced by $N_y$ flux quanta and its fundamental domain is a rectangle of dimensions $L_x$ by $L_y$.
We therefore have $L_x L_y / 2 \pi l^2 = N_y$.
Since there is a one-to-one correspondence between $\ket{K_y}$ and $\ket{\mathcal{K}_y}$, we will write both of them as $\ket{K_y}$ in the following. It will be clear from the context whether the lattice or the continuum state is referred to.

For the sake of clarity, we first treat the case of  the contact interaction ($m=0$).
In this case, the interaction is uniform in momentum space: $V^{(0)}(({\bf q}+{\bf H})^2 l^2)=1$.
We can therefore identify $F_{\text{LLL}}^{(0)} ({\bf q})$ with the continuum counterpart of the lattice form factor $d_0(\bq)$.
This form factor, which we will call $d_\text{LLL}({\bf q})$, is given by a periodicized Gaussian:
\begin{equation}
\begin{aligned}
|d_\text{LLL}({\bf q})|^2 & \equiv F_{\text{LLL}}^{(0)} ({\bf q}) \\
&= \sum_{{\bf H}}  e^{- \frac12 ({\bf q}+{\bf H})^2 l^2}  \\
& = \sum_{H_x, H_y \in \mathbb{Z}}  e^{- \frac12 \frac{1}{N_y} 2 \pi \left[ \tau (q_x + H_x N_y)^2 + \frac1{\tau} (q_y + H_y N_y)^2 \right]} 
\end{aligned}
\end{equation}
where $\tau = L_y / L_x$.

\begin{figure}
  \centering
  \begin{tabular}{@{}p{0.45\linewidth}@{\quad}p{0.45\linewidth}@{}}
    \subfigimg[width=\linewidth]{a)}{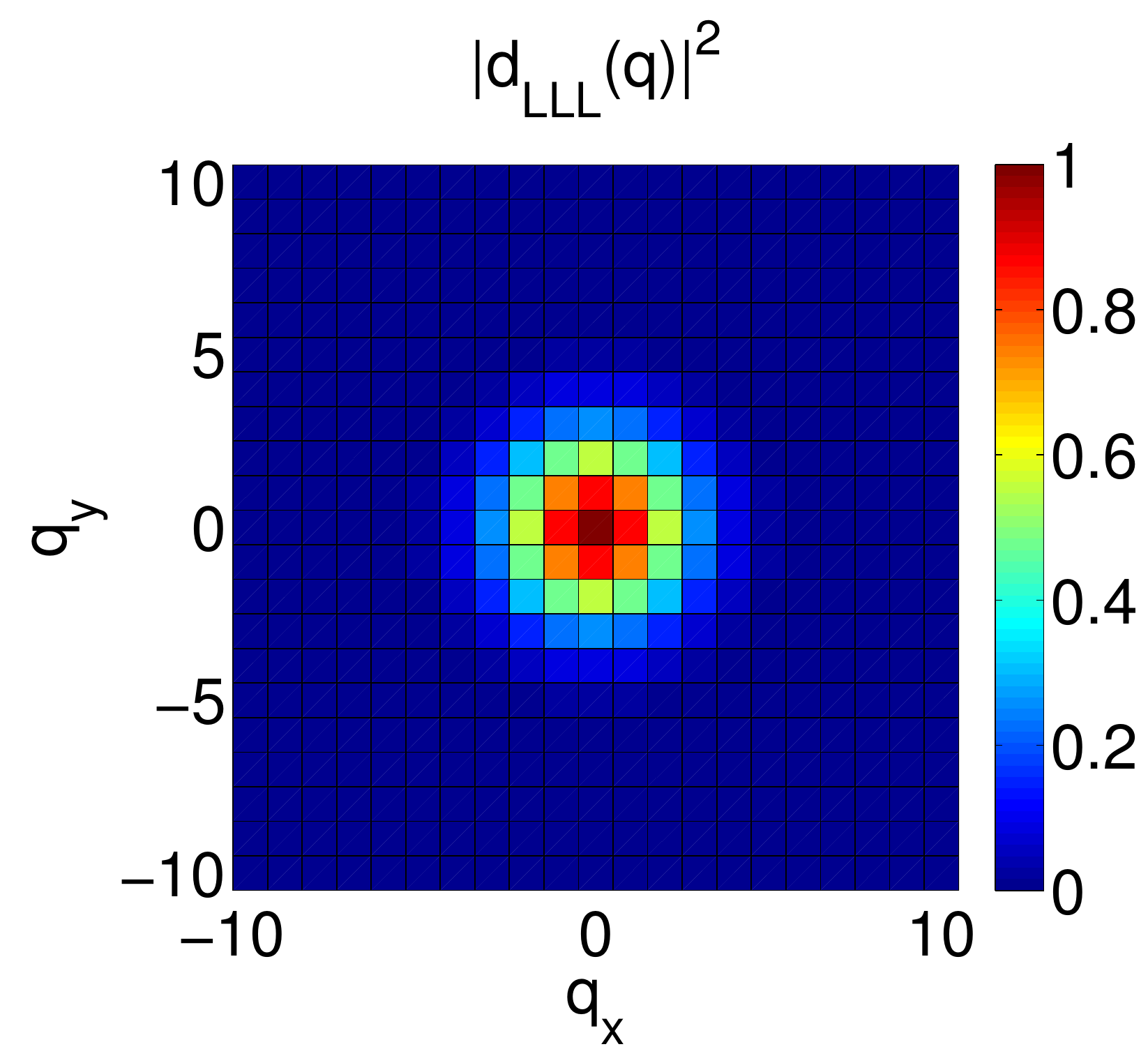} &
    \subfigimg[width=\linewidth]{b)}{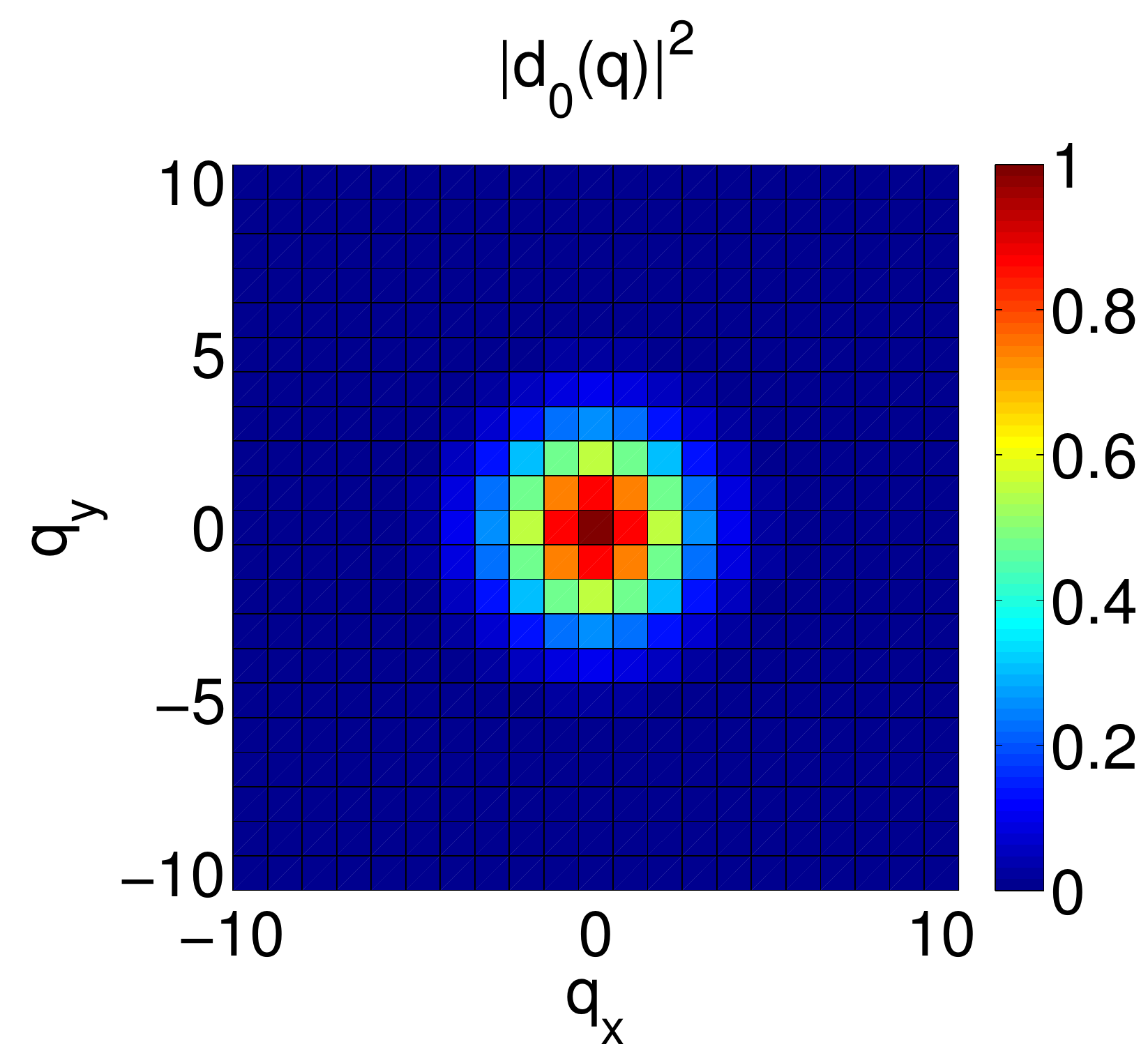} \\
    \subfigimg[width=\linewidth]{c)}{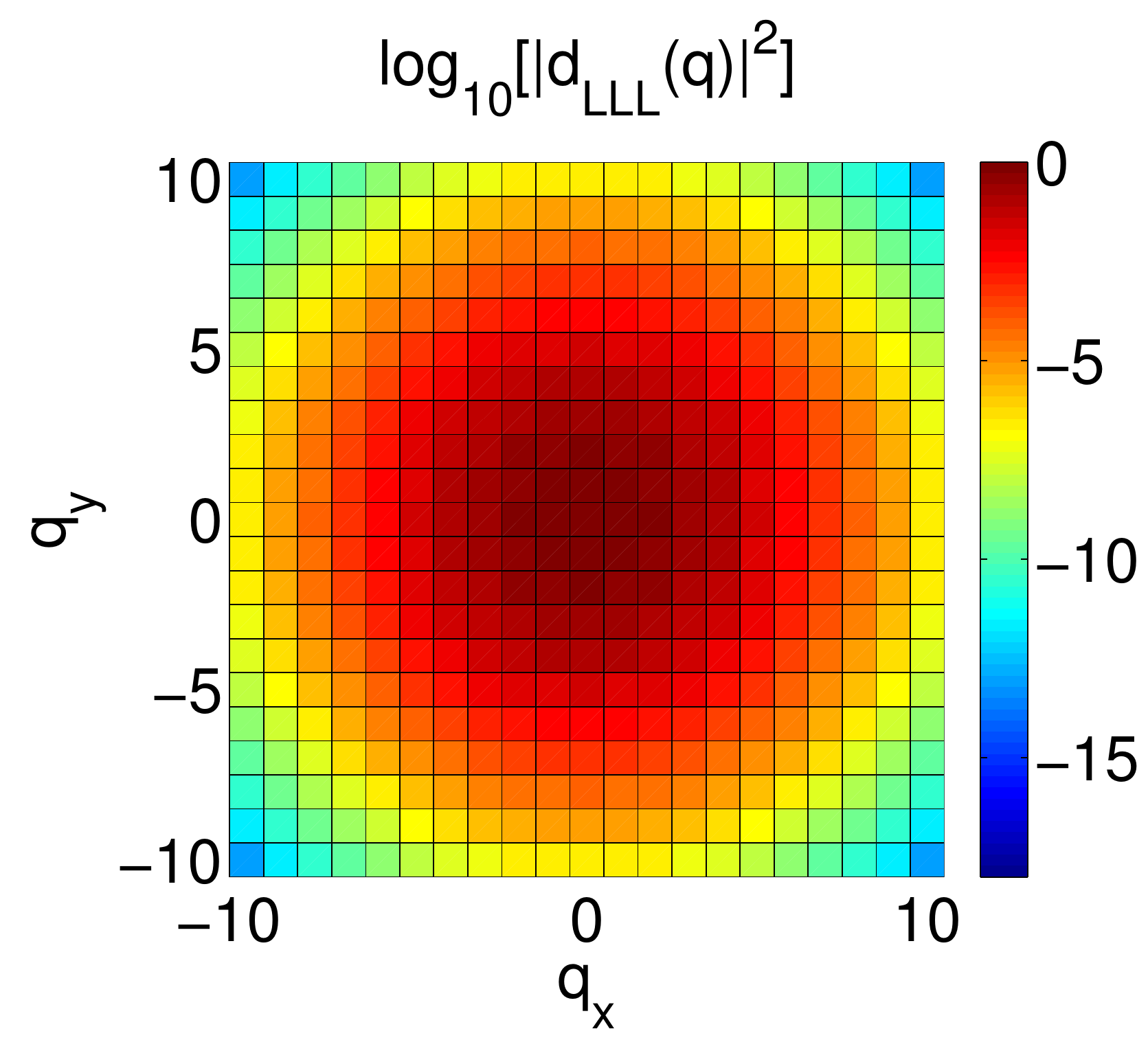} &
    \subfigimg[width=\linewidth]{d)}{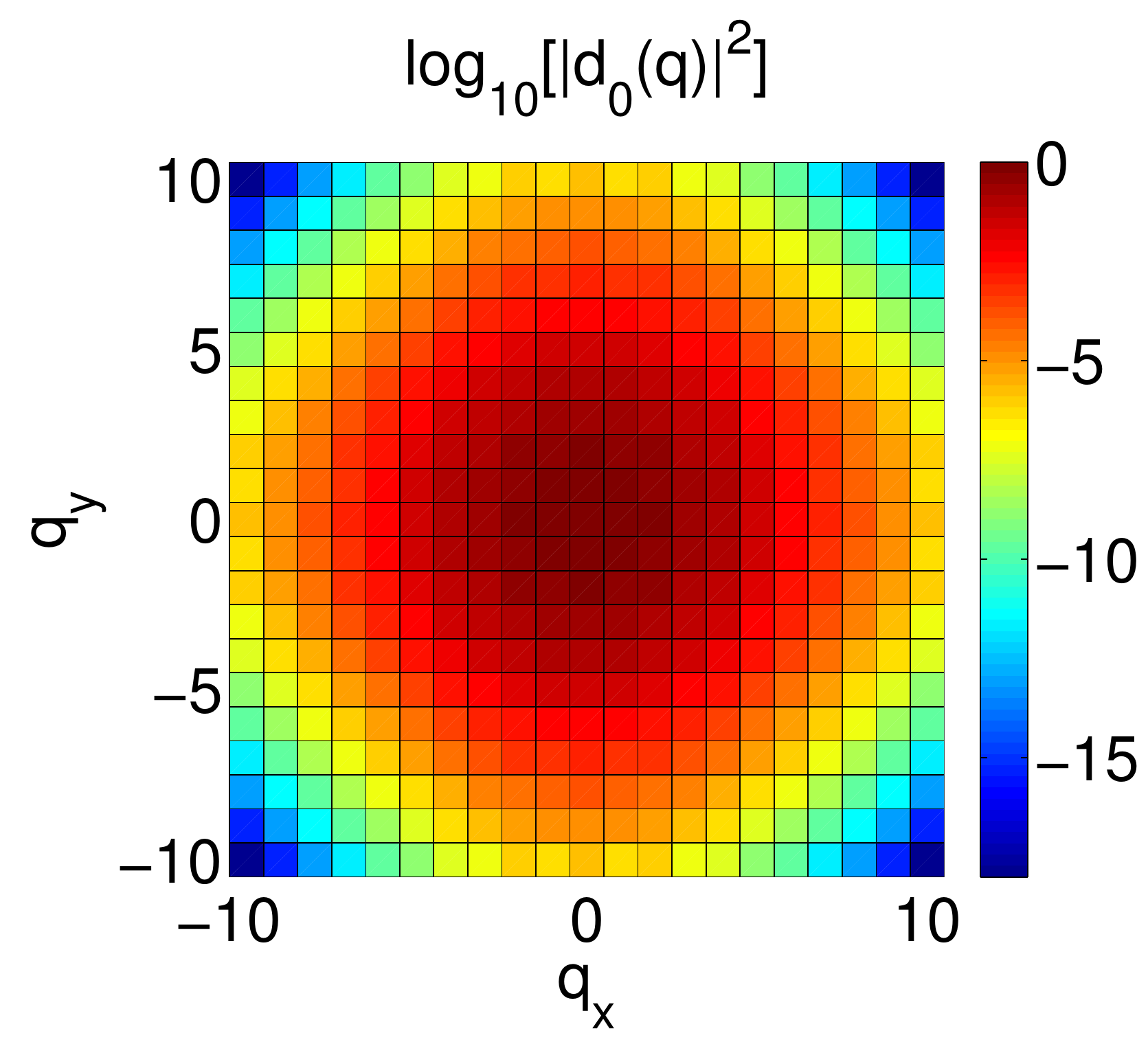} 
  \end{tabular}
\caption{\label{FqC1}
Form factors in the LLL and the lattice case. The form factor in the lattice case is given for nearest-neighbor hopping. The system size is given by $N_y=21$.
}
\end{figure}

By comparing Eqs. (\ref{PHP1}) and (\ref{PHPc1}), it is clear that we have reduced the difference between the interacting Hofstadter model and interacting electrons in a Landau level to a very simple thing: the difference between two real scalar functions defined on a discrete $N_y$ by $N_y$ BZ: $|d_0({\bf q})|^2$ and $|d_\text{LLL}({\bf q})|^2$. 
These two functions are plotted in Fig.~\ref{FqC1}.

Since there is no closed analytical formula for $|d_0({\bf q})|^2$, it was computed numerically. 
The continuum form factor $|d_\text{LLL}({\bf q})|^2$ has only one parameter $\tau$, which makes the Gaussian elongated in the $y$ direction when chosen larger than one.
As explained before, $|d_0({\bf q})|^2$ contains the only information remaining from the cyclotron degrees of freedom and therefore depends on the single-particle hoppings present in $H_0$ ($t_x$ and $t_y$ in the case of nearest-neighbor hopping).
We find numerically that $|d_0({\bf q})|^2$ is very close to a Gaussian, up to lattice effects of order $1/Q$.
It is $x \leftrightarrow y$ symmetric when $t_x=t_y$, and becomes elongated in the $y$ direction when $t_y > t_x$.


For the sake of simplicity, we treat the $x \leftrightarrow y$ symmetric case on both counts ($\tau=1$ and $t_y/t_x=1$).
As seen in Fig.~\ref{FqC1}, the difference between $|d_0({\bf q})|^2$ and $|d_\text{LLL}({\bf q})|^2$ is small.
Furthermore, in agreement with Ref.~\cite{PhysRevB.90.075104}, we find that it decays like $1/Q$:
\begin{equation}
 |d_0({\bf q})|^2 -  |d_\text{LLL}({\bf q})|^2 \propto \frac1{Q} \ll 1.
\end{equation}
As expected, the difference vanishes in the small flux density limit $Q \rightarrow \infty$ in which limit it was shown analytically\cite{PhysRevB.90.075104} and numerically\cite{PhysRevLett.94.086803, PhysRevA.76.023613} that the Hofstadter model tends to the continuum case.

\begin{figure}
  \centering
  \begin{tabular}{@{}p{0.45\linewidth}@{\quad}p{0.45\linewidth}@{}}
    \subfigimg[width=\linewidth]{a)}{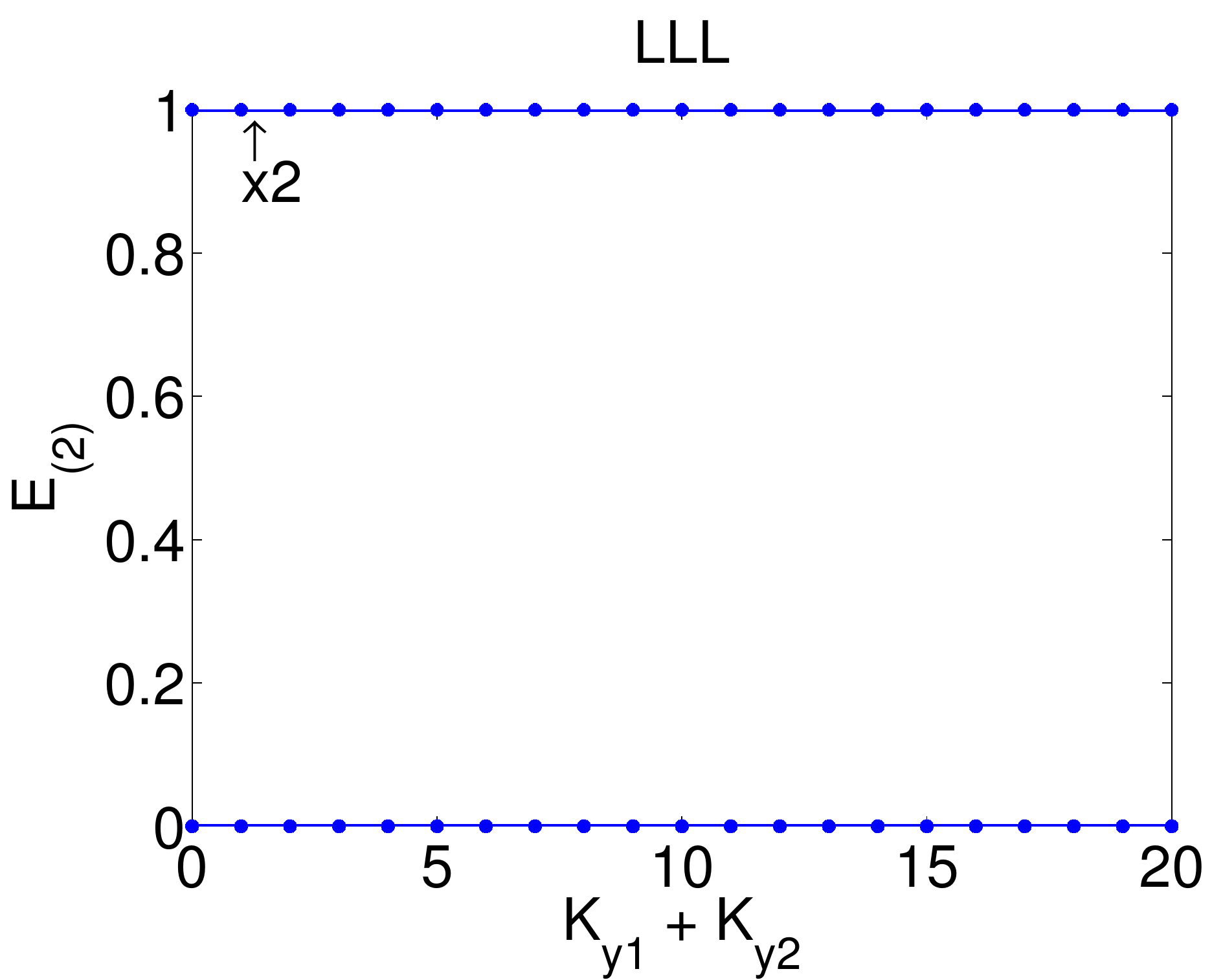} &
    \subfigimg[width=\linewidth]{b)}{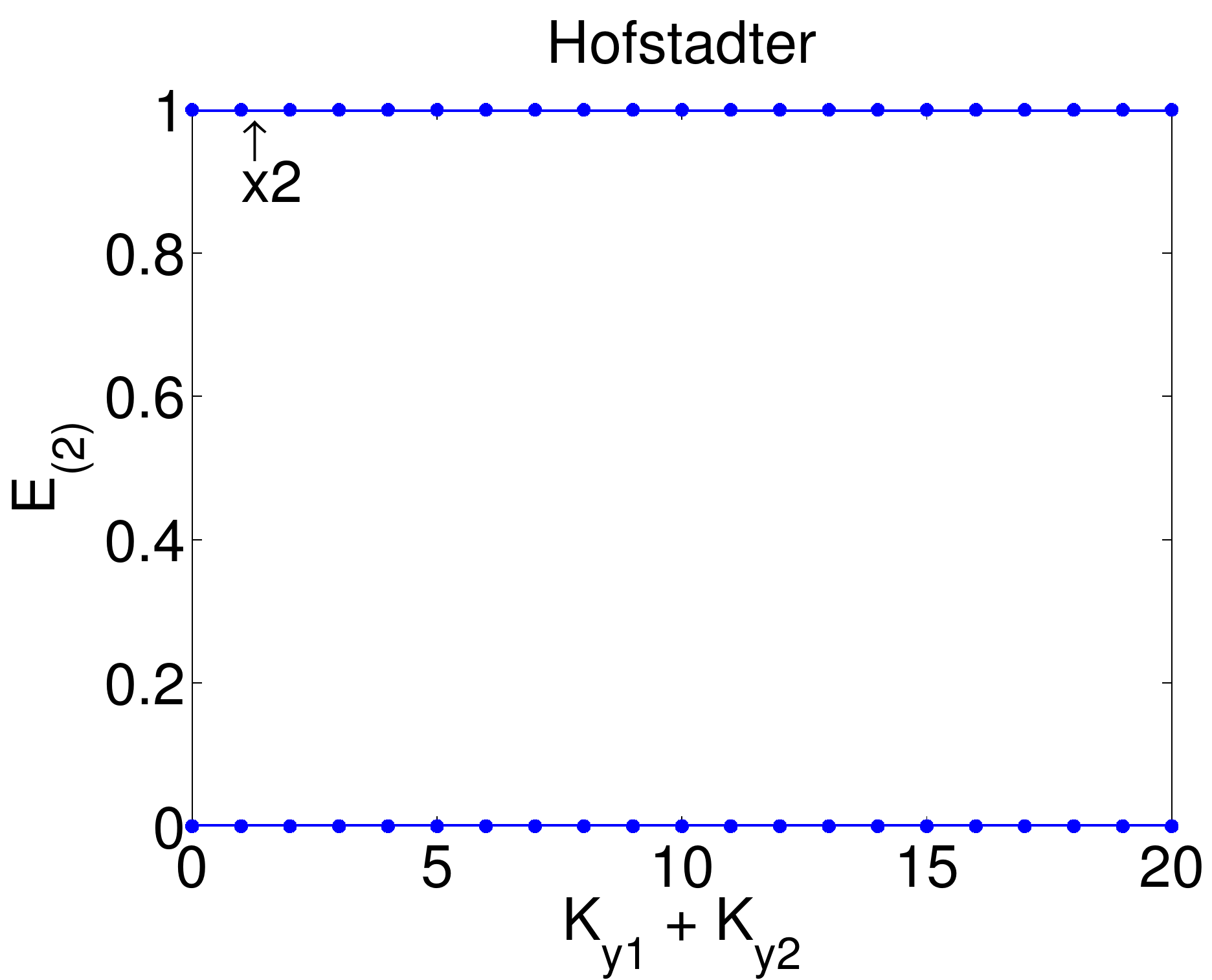} \\
    \subfigimg[width=\linewidth]{c)}{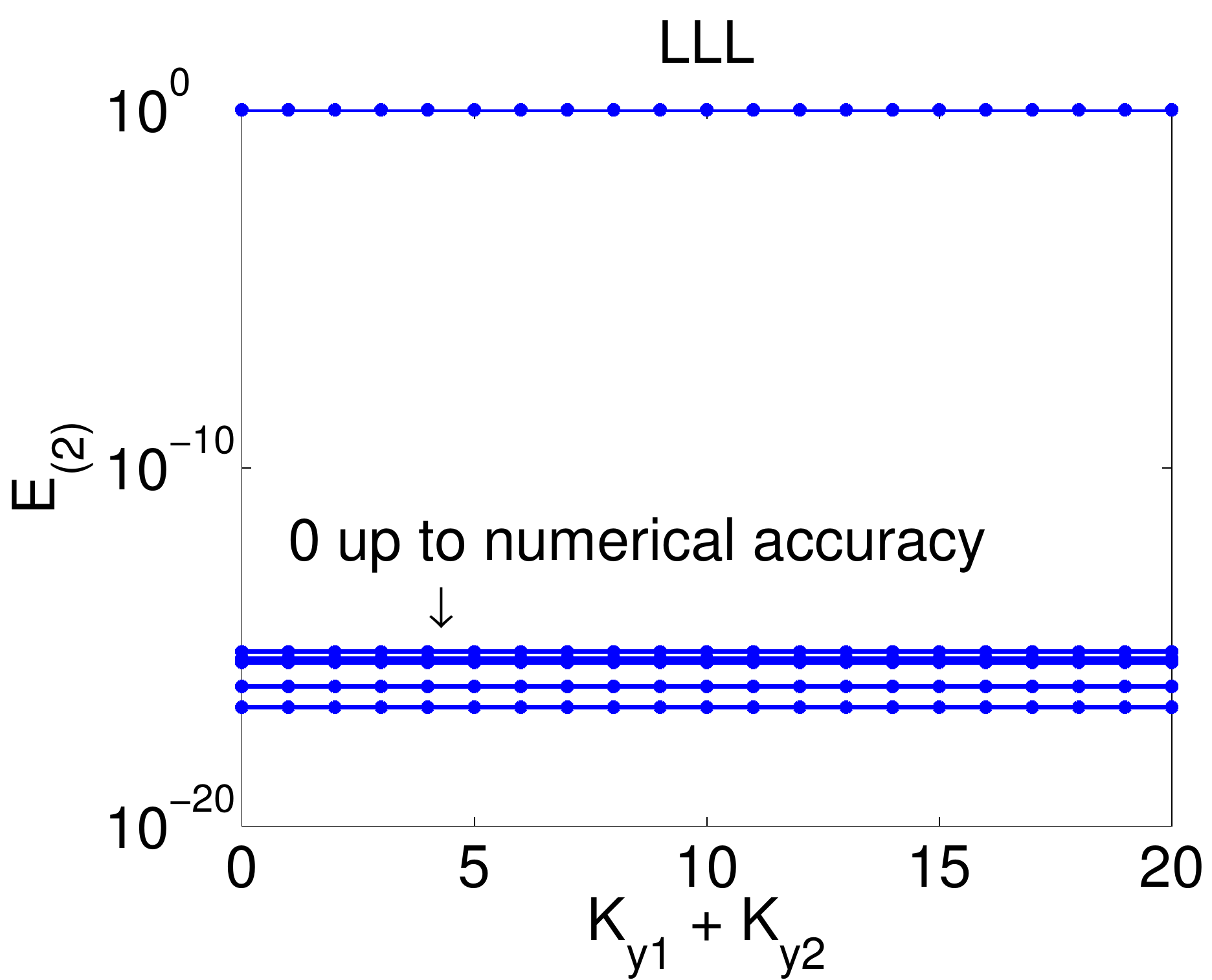} &
    \subfigimg[width=\linewidth]{d)}{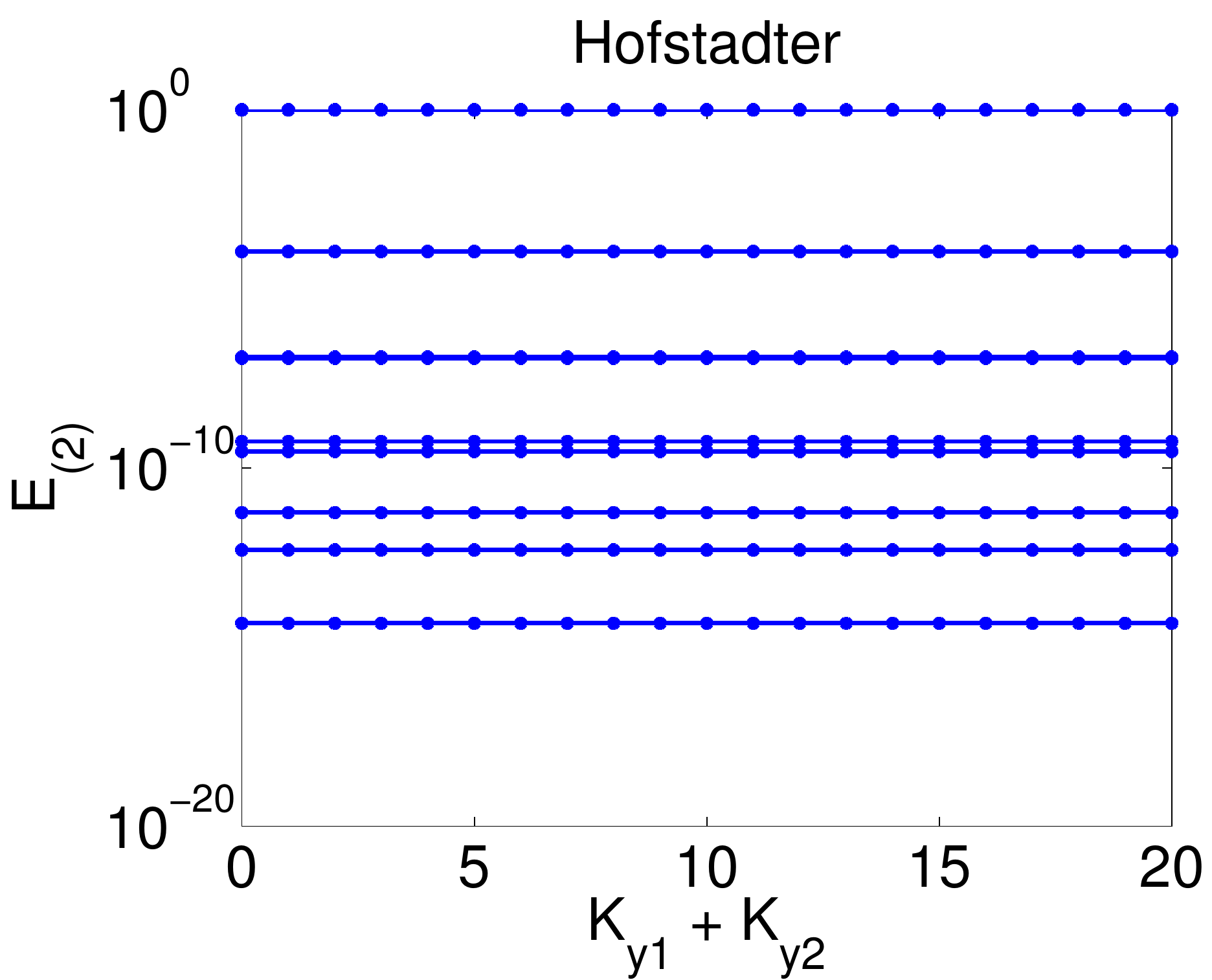}
  \end{tabular}
\caption{\label{TwoBodySpec}
Two-particle spectrum $E_{(2)}$ of the interaction Hamiltonian vs momentum sector $K_{y1} + K_{y2}$ for a contact interaction projected to the continuum LLL (left) and to the lowest band of the Hofstadter model with our special system size (right). 
These spectra were obtained by exact diagonalization for $N_y=21$. 
As indicated in the top panels, there are two nearly degenerate levels close to $E=1$ corresponding to the 0-th pseudopotential\cite{PhysRevLett.111.126802}. 
The bottom panels show the same spectrum on a logarithmic scale, revealing the differences between the LLL and the Hofstadter case (see main text).
%
}
\end{figure}

It is well known that $\bar{H}_c$ has an exact zero-energy ground state given by the $\nu=1/2$ bosonic Laughlin state\cite{PhysRevB.31.2529}.
%
Since, for finite $Q$, $|d_0({\bf q})|^2$ and $ |d_\text{LLL}({\bf q})|^2$ are close but strictly different, the FQH-like ground state of the Hofstadter model with a contact interaction, also known as the Hofstadter-Hubbard model \cite{PhysRevLett.104.145301}, will in general be at small but finite energy, and its overlap with the $\nu=1/2$ Laughlin state will be close to, but not exactly one.

One simple way to quantify this small difference between the two models is to look at the two-particle spectra. 
They have been shown to give valuable information in terms of a pseudopotential analogy about the possible FQH-like phases that appear in a given lattice model\cite{PhysRevLett.111.126802, PhysRevB.88.205101}. 
In this analogy, each pair of nearly degenerate levels at non zero energy $E$ corresponds to a different pseudopotential of index $m$ with a coefficient $\nu_m = E$.
For example, in the LLL case with a contact interaction, there is only one of such pairs, corresponding to $m=0$ (see left of Fig.~\ref{TwoBodySpec}).
On the contrary, in the Hofstadter case (see right panels of Fig.~\ref{TwoBodySpec}), there are additional levels in the spectrum at much smaller energy, which can be interpreted as revealing the presence of higher pseudopotentials ($m>0$) with much smaller coefficients $\nu_m$.
These higher pseudopotentials are the reason why the overlap between the ground state of the Hofstadter-Hubbard model and the $\nu=1/2$ Laughlin state is close to, but not exactly one.

While for pseudopotentials in a LL, two-particle energy levels have exactly zero dispersion with momentum sector $K_{y1}+K_{y2}$, this is not the case in general for lattice models.
This finite dispersion present in lattice models has been conjectured to destabilize FQH phases\cite{ PhysRevB.88.205101}.
As can be seen in Fig~\ref{TwoBodySpec}, for the Hofstadter model with our system size, this dispersion exactly vanishes, like for Landau levels.
We should also mention that the topological center-of-mass degeneracy of the FQH states on the torus is exact for our system size, while it is only approximate in general for lattice systems. 
Based on the $T_{x,y}$ operators defined in Eq.~(\ref{TOp}), one can indeed construct a center-of-mass magnetic translation operator that commutes with the interaction, just like in the continuum case.

Instead of using an on-site interaction on the lattice and looking at the difference between  $\bar{H}_{\text{Lat}}$ and $\bar{H}_c$, one can tune the interaction on the lattice so as to have $\bar{H}_{\text{Lat}} = \bar{H}_c$. 
For the sake of clarity, we should emphasize that this equality means that the two projected Hamiltonians have exactly the same matrix elements in the occupation number basis of their respective $\ket{K_y}$ states. 
Since $F_{\text{Lat}}({\bf q}) = V({\bf q}) |d_0({\bf q})|^2$, one simply has to take
\begin{equation}
V({\bf q}) = \frac{ |d_\text{LLL}({\bf q})|^2}{|d_0({\bf q})|^2}\text{.}
\label{VqEq}
\end{equation}
This formula will obviously break down if $|d_0({\bf q})|^2$ vanishes for some $\bq$.
We find that $|d_0(N_y/2,N_y/2)|^2=0$ for nearest-neighbor hopping, but that $|d_0(\bq)|^2 \neq 0$ for all $\bq$ as soon as next to nearest neighbor(NNN) hopping is added. 
This formula can therefore be applied to the Hofstadter model whose hopping range is at least NNN to find the interaction on the lattice that will give, after projection, any desired pseudopotential Hamiltonian. 

\begin{figure}
  \centering
  \begin{tabular}{@{}p{0.45\linewidth}@{\quad}p{0.45\linewidth}@{}}
    \subfigimg[width=\linewidth]{a)}{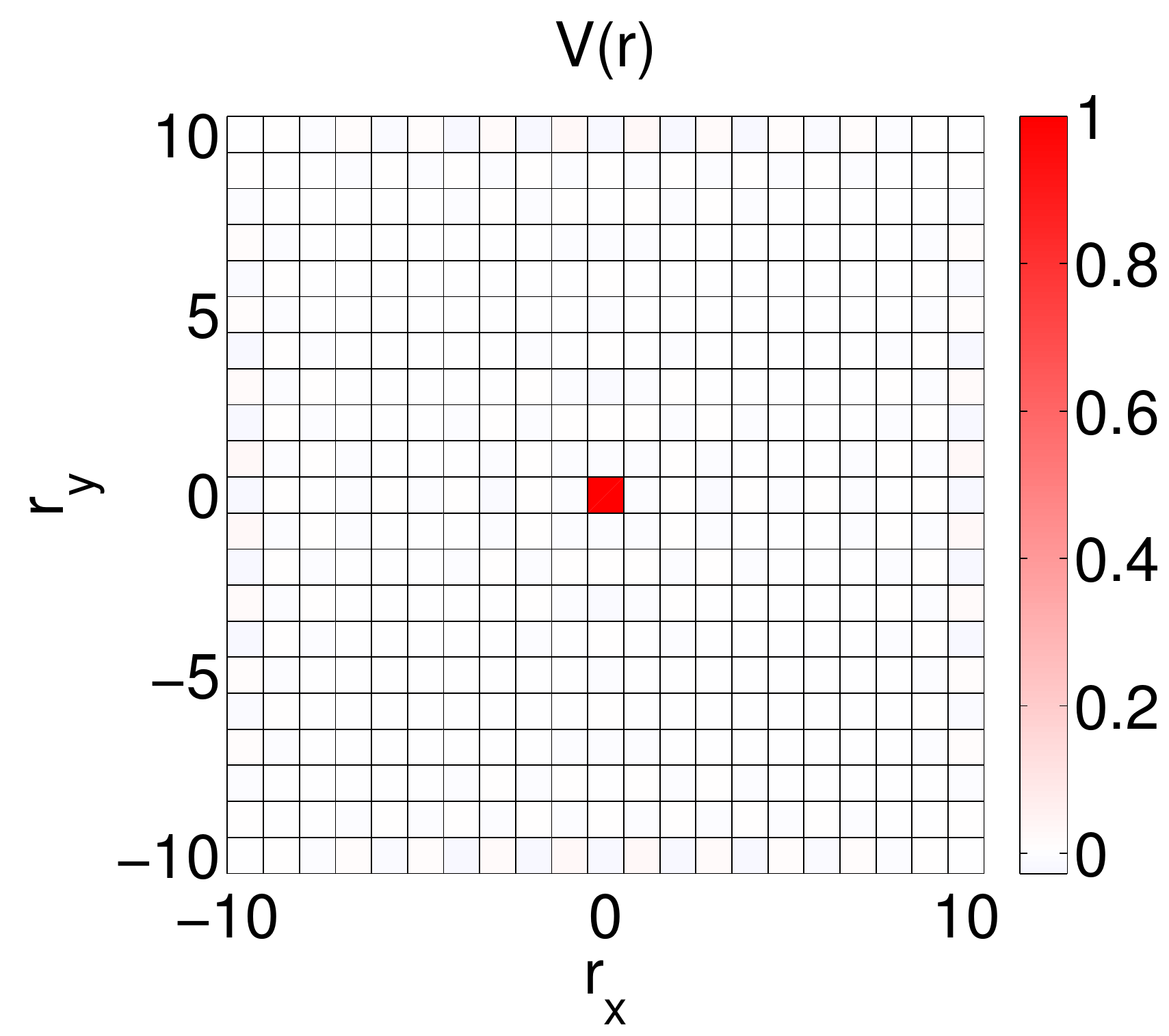} &
    \subfigimg[width=\linewidth]{b)}{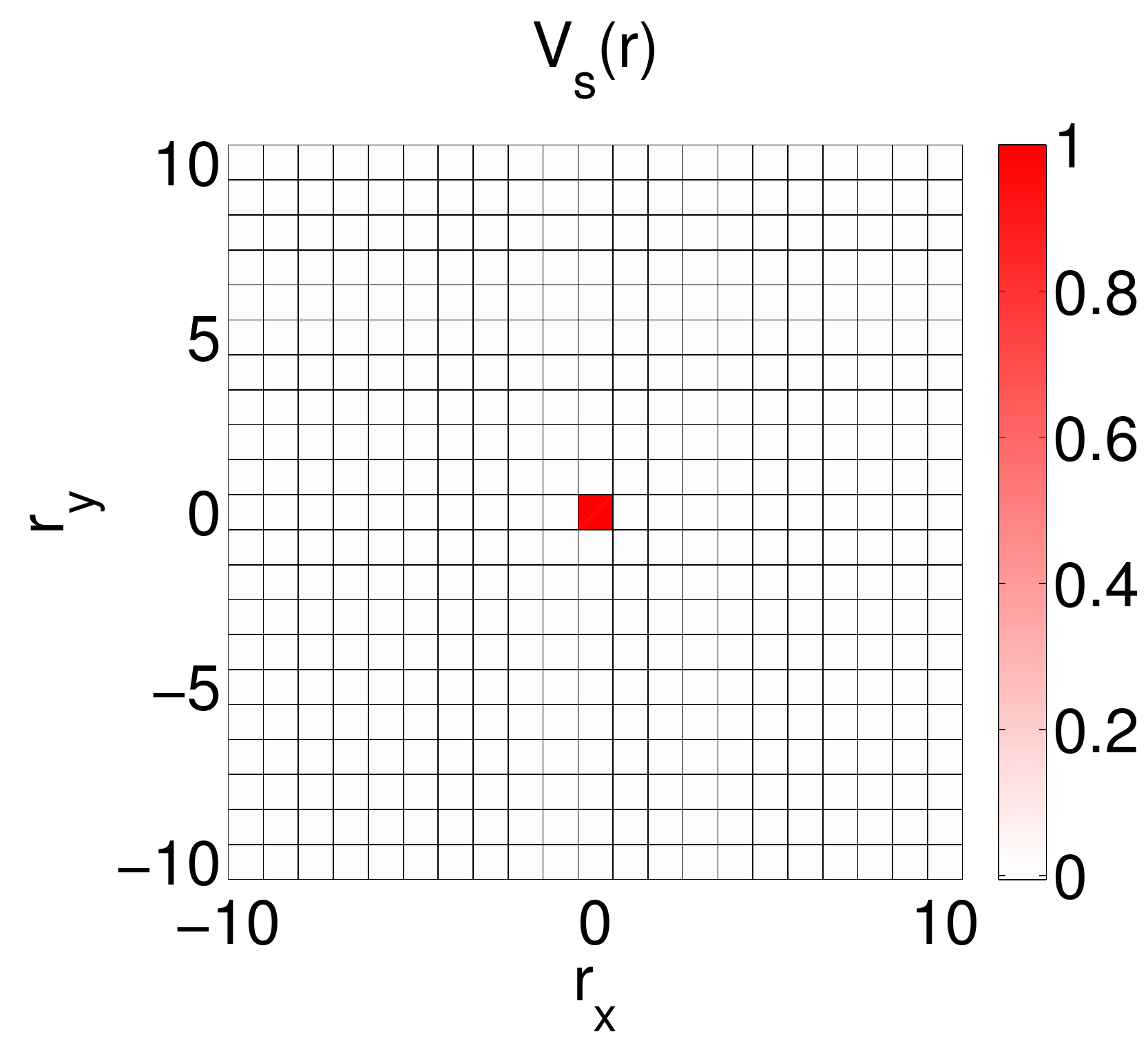} \\
    \subfigimg[width=\linewidth]{c)}{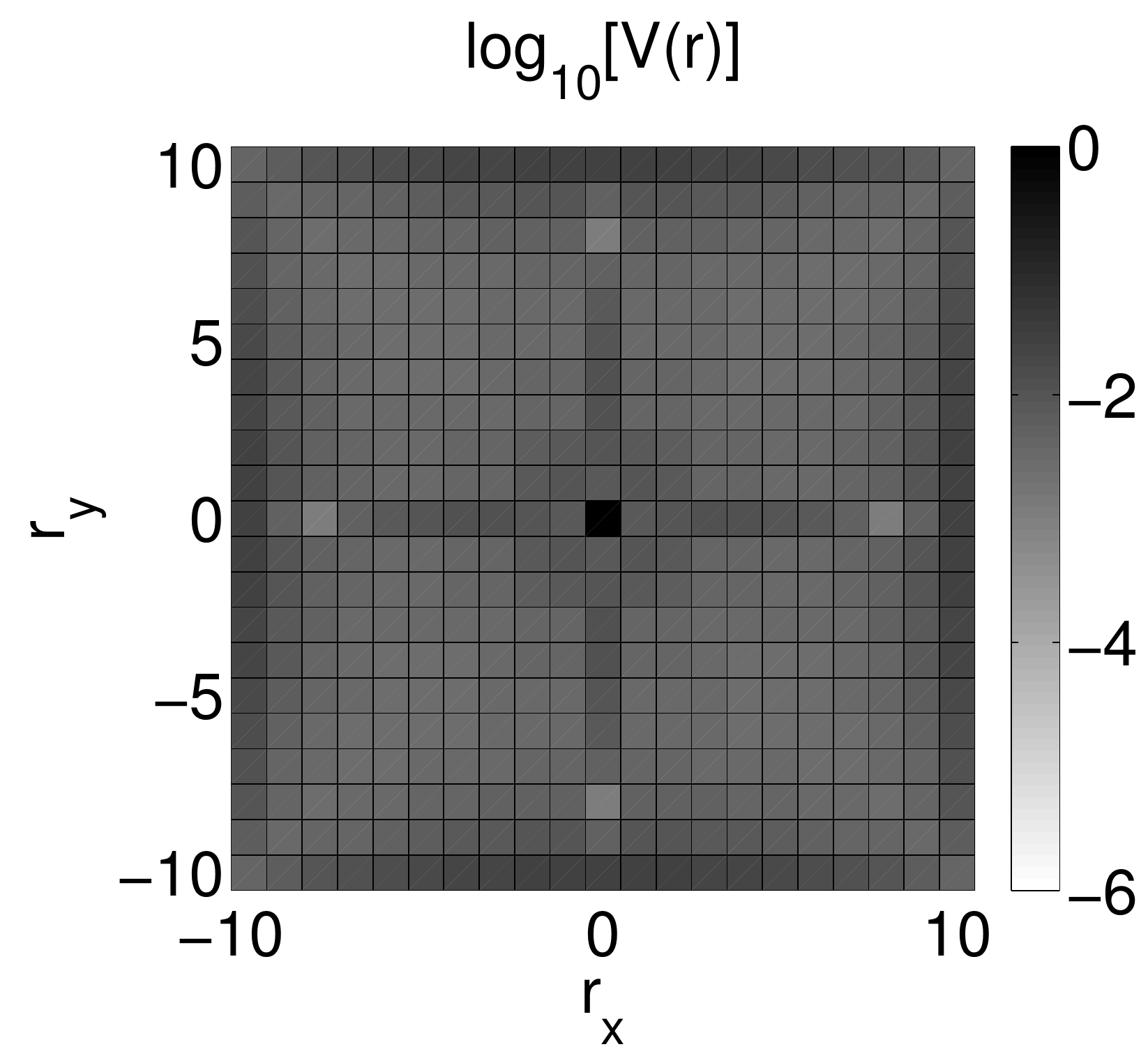} &
    \subfigimg[width=\linewidth]{d)}{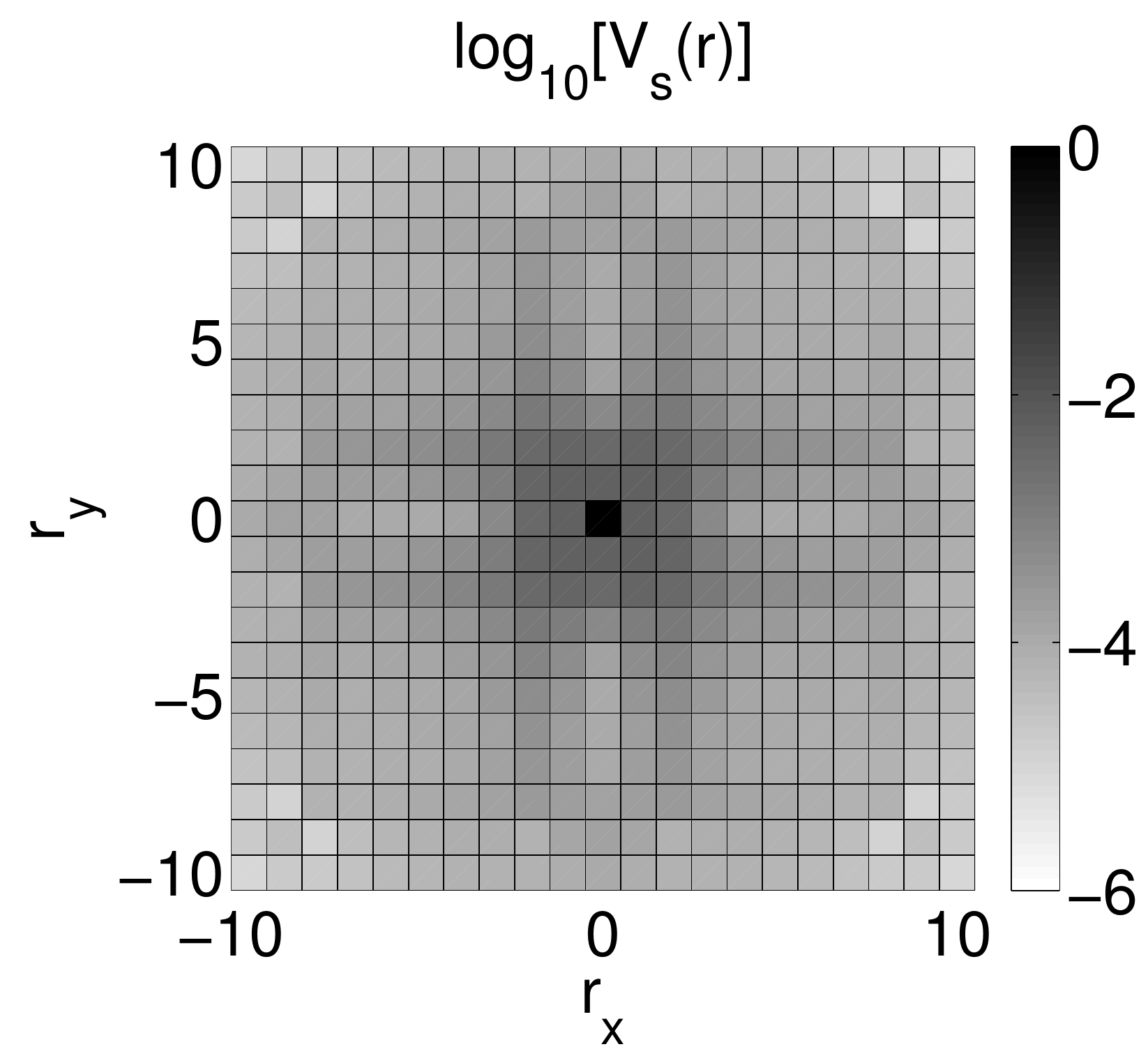}
  \end{tabular}
\caption{\label{Vq}
Interaction potential for the Hofstadter model which leads to a Hamiltonian with an exact zero-energy ground state given by the $\nu=1/2$ continuum Laughlin ground state. 
The left panels show $V(\br)$, the interaction leading to the LLL form factor, while the right panels show $V_s(\br)$, the interaction leading to the form factor of the Kapit-Mueller model\cite{PhysRevLett.105.215303}.
The nearest neighbor hopping is 1 and the NNN hopping is 0.27. 
The plots are given for $N_y = 21$. 
The on-site interaction is normalized to 1. 
The second largest interaction is on the NNN and is equal to 0.029 for $V(\br)$ and 0.027 for $V_s(\br)$.
}
\end{figure}

We show a typical plot of this interaction in real space $V(\br)$ in Figs.~\ref{Vq}(a) and \ref{Vq}(c) for the case of the 0-th pseudopotential.
This interaction is a short-range peak with a long-range small amplitude component.

There is actually a way to heavily suppress this long-range part.
Kapit and Mueller proposed a hopping model\cite{PhysRevLett.105.215303} for which the single-particle wave functions are the continuum ones sampled at the lattice points and for which they showed that the $\nu=1/2$ continuum Laughlin state sampled at the lattice points is the exact zero-energy ground state in the case of an on-site interaction.
As explained in the next section, the Kapit-Mueller model leads to a form factor different than that of the LLL but for which the Laughlin wave function is still the exact ground state.
If we call this form factor $|d_s({\bf q})|^2$, we can compute another interaction $V_s(\bq)$ defined as
\begin{equation}
V_s({\bf q}) = \frac{|d_s({\bf q})|^2}{|d_0({\bf q})|^2}.
\end{equation}
As shown in Figs.~\ref{Vq}(b) and \ref{Vq}(d), this interaction is better behaved than $V(\bq)$: the long-range oscillations are heavily suppressed.

%

Finally, we should emphasize that the condition $N_x=1, N_y=Q$ (which corresponds to a square system of $Q$ by $Q$ plaquettes) allows us to reach arbitrarily large system sizes, at the expense of taking a large value for $Q$. 
If we keep $P$ fixed at 1, the large $Q$ limit corresponds to $n_{\phi} \rightarrow 0$, in which case Landau levels are recovered.

An arbitrarily large system size can also be obtained without having $n_{\phi}$ tending to zero by taking the following limit for a given $C>1$:
\begin{equation}
\begin{aligned}
Q, P &\rightarrow \infty, \\
 P/Q &\rightarrow 1/C, \\
 P \ C &\mod Q = 1.
\end{aligned}
\end{equation}
This limit can be implemented in practice by taking $P \in \mathbb{Z}$ and $Q=PC-1$.
It is easy to see that $P$ and $Q$ are always coprime and that $P/Q$ tends to $1/C$ when $P \rightarrow \infty$, as required.
In this case, one obtains bands of Chern number $C$ with exactly flat dispersion and Berry curvature. 
The Hofstadter model for flux densities close to rational fractions $1/C$ has already been studied in previous work\cite{PhysRevLett.96.180407,PhysRevA.78.013609,PhysRevLett.108.256809, PhysRevB.90.075104}.
We will discuss this case in Sec. IV A.



\section{Relation to previous work}
\subsection{Wannier construction}
In a previous work, Qi showed that one can obtain an interaction for any TFB in a lattice model that leads to the same Hamiltonian as the continuum pseudopotentials\cite{PhysRevLett.107.126803,PhysRevB.88.035101,PhysRevB.90.085103}.
This construction works as follows. 
The lattice single-particle basis inside one TFB can be chosen to be the Wannier states $\ket{K_y}$, with $K_y=0, \dots, N_{\phi}-1$. 
There is then a one-to-one correspondence with the single-particle basis of the LLL in the continuum case. 
The lattice interaction Hamiltonian is chosen to have the same matrix elements as the desired projected continuum interaction Hamiltonian in the occupation number basis of the $\ket{K_y}$ states:
\begin{equation}
\bar{H} = \sum_{K_{y1} + K_{y2} = K_{y3} + K_{y4}} V_{K_{y1},K_{y2},K_{y3},K_{y4}} c^{\dagger}_{K_{y1}} c^{\dagger}_{K_{y2}} c_{K_{y3}} c_{K_{y4}}, 
\end{equation}
where $V_{K_{y1},K_{y2},K_{y3},K_{y4}}$ is the same as for the desired continuum pseudopotential Hamiltonian.
One can then find the corresponding lattice interaction in real space $V(\br_1, \br_2, \br_3, \br_4)$ by replacing $c_{K_{y}}$ by $\sum_{\br} \psi_{K_y}(\br) c_{\br}$ where $ \psi_{K_y}(\br)$ is the Wannier wavefunction.
This interaction will in general be of the range of the magnetic length (more precisely, its lattice counterpart), and will not be density-density (there will be ring exchange terms).
The reason for this is that $V(\br_1, \br_2, \br_3, \br_4)$ corresponds to the projected interaction $\bar{H}$ and not the bare interaction before projection.
One could add to $V$ any interaction $V'$ that projects to zero and still get the same desired projected Hamiltonian.
By choosing $V'$ appropriately, it should be in general possible to minimize the range of the interaction and the ring exchange terms.
Nevertheless, finding $V'$ in general is a hard problem and is model-specific.

On the contrary, our construction provides a way of finding directly the unprojected interaction [defined in Eq.~(\ref{VqEq})] that is only density-density and of minimal range and that leads to a desired continuum pseudopotential Hamiltonian after projection; the drawback being that it only works for the Hofstadter model at certain system sizes.



\subsection{Kapit-Mueller model}
In brief, we obtained exactly flat bands and Berry curvature in the Hofstdter model for any hopping (including strictly short range hopping) by choosing the right system size for the right flux density.
This special system size also enabled us to find density-density interactions on the lattice that lead to the exact same Hamiltonian as the pseudopotentials in the continuum.
In this section, we will relate this result to the previous work of Kapit and Mueller\cite{PhysRevLett.105.215303}(see also Ref.~\cite{PhysRevA.88.033612}), who showed that, with the appropriate hoppings, the Hofstadter model can give rise to an exactly flat lowest energy band whose single-particle wave functions are the continuum LLL wave functions sampled at the lattice points.
They could then use the same argument as in the continuum to find an exact zero-energy eigenstate of an on-site interaction projected to the lowest band, thereby providing a lattice equivalent of the $\nu=1/2$ Laughlin state.
The hoppings they use decay exponentially with distance.
The Kapit-Mueller (KM) model works for any system size and flux density, unlike ours, but we will show that the case $N_y=Q, N_x=1$ is still a peculiar system size; it is the only system size for which the lattice Laughlin state is exactly the same as the continuum one, in the sense that they have the same many-body wave function in the occupation number basis of the single-particle states $\ket{\mathbf{K}}$.

The continuum LLL wavefunctions on the torus are given by 
\begin{equation} 
\begin{aligned}
&\psi^c_{\mathbf{K}}(\tilde{x}, \tilde{y}) = \sum_{n \in \mathbb{Z}} \sum_{m=0}^{N_x-1} \exp\left[ i \frac{2 \pi}{L_y} (K_y+ m N_y + n N_{\phi}) \tilde{y}  \right] \times \\
& \exp\left[i 2 \pi m \frac{K_x}{N_x}\right] \exp\left[ -\frac{\left(\tilde{x} + l^2 (K_y + m N_y + n N_{\phi}) \frac{2 \pi}{L_y}\right)^2  }{2 l^2} \right] 
\label{cont}
\end{aligned}
\end{equation}
where $K_{x,y} = 0, \dots, N_{x,y}-1$ and $N_{\phi}=N_x N_y$. 
Note that this equation is simply the generalization of Eq.~\ref{OneBodyStatesKy} to the case of $N_x \neq 1$.
The lattice version of these states is obtained by imposing $P$ fluxes per unit cell ($L_x L_y / 2 \pi l^2 = P N_{\phi}$).
In the following, we will restrict ourselves to the case of $P=1$ for the sake of clarity.
The sampling is given by $\tilde{x}=x L_x/N_x Q$ and $\tilde{y}=y L_y/N_y$ for a lattice of $N_x Q$ by $N_y$ plaquettes.
The sampled wavefunctions are derived in Appendix A and are given by
\begin{equation} 
\begin{aligned}
&\psi^s_{\mathbf{K}}(x, y) = \sum_{m \in \mathbb{Z}} \exp\left[ i \frac{2 \pi}{N_y} K_y y  \right] \exp\left[i 2 \pi m \frac{K_x}{N_x}\right] \times\\
& \exp\left[ -\frac12 2 \pi \frac{1}{Q} \left(x + \frac1{N_y} Q (K_y + m N_y ) \right)^2   \right]. 
\label{sampled}
\end{aligned}
\end{equation}

The $\nu=1/2$ Laughlin wavefunction has a well-known expression in terms of theta functions\cite{PhysRevB.31.2529}. Since it lies only in the LLL, it can generically be written as a sum of products of single-particle wavefunctions:
\begin{equation}
\Psi^{c}(\tilde{z}_1\dots \tilde{z}_n) =  \sum_{\mathbf{K}_1 \dots \mathbf{K}_N} \phi^c(\mathbf{K}_1 \dots \mathbf{K}_N)  \prod_{i=1}^N \psi^c_{\mathbf{K}_i}(\tilde{z}_i)
\end{equation}
where $\tilde{z} = \tilde{x} + i \tilde{y}$.
Kapit and Mueller's lattice many-body wavefunction is the one obtained by replacing $\tilde{z}$ in the continuum Laughlin wavefunction by its lattice equivalent $z=x+iy$\cite{PhysRevLett.105.215303}. It can therefore be written as
\begin{equation}
\Psi^{s}(z_1\dots z_n) =  \sum_{\mathbf{K}_1 \dots \mathbf{K}_N} \phi^c(\mathbf{K}_1 \dots \mathbf{K}_N)  \prod_{i=1}^N \psi^s_{\mathbf{K}_i}(z_i)
\end{equation}
where it should be emphasized that it is the same $\phi^c(\mathbf{K}_1 \dots \mathbf{K}_N) $ as in the continuum.
The difference with the continuum is that the sampled wavefunctions $\psi^s_{K_{y}}(z)$ do not form in general an orthonormal basis. 
After a change of basis, we obtain
\begin{equation}
\Psi^{s}(\mathbf{K}_1 \dots \mathbf{K}_N) =  \sum_{\mathbf{K}'_1 \dots \mathbf{K}'_N} \phi^c(\mathbf{K}'_1 \dots \mathbf{K}'_N)  \prod_{i=1}^N \sand{\psi^s_{\mathbf{K}_i}}{\psi^s_{\mathbf{K}'_i}}
\end{equation}
where $\sand{\psi^s_{\mathbf{K}_i}}{\psi^s_{\mathbf{K}'_i}} = \sum_z (\psi^{s}_{\mathbf{K}_i}(z))^* \psi^{s}_{\mathbf{K}'_i}(z)$. 

These overlaps are calculated in Appendix B. 
From them, we see that the sampled wavefunctions are orthogonal but their norm is in general $\mathbf{K}$-dependent:
\begin{equation}
\sand{\psi^s_{\mathbf{K}_i}}{\psi^s_{\mathbf{K}'_i}} = \delta_{\mathbf{K}_i,\mathbf{K}'_i} \mathcal{N}(\mathbf{K}_i)^2. 
\end{equation}
Since $\Psi^{c}(\mathbf{K}_1 \dots \mathbf{K}_N) = \phi^c(\mathbf{K}_1 \dots \mathbf{K}_N)$ by construction, we obtain in this case
\begin{equation}
\Psi^{s}(\mathbf{K}_1 \dots \mathbf{K}_N) =  \Psi^{c}(\mathbf{K}_1 \dots \mathbf{K}_N)  \prod_{i=1}^N \mathcal{N}(\mathbf{K}_i)^2.
\end{equation}

Interestingly, if $\mathcal{N}(\mathbf{K}_i) =  \mathcal{N}$ for all $\mathbf{K}$, the two wavefunctions are equal up to an overall multiplicative constant. 
Based on the calculation of the overlaps given in Appendix B, we see that this is the case only when $Q$ is a multiple of $N_y$ and when $ N_x=1$, i.e. for our special system size.

This means that, for our system size, the continuum and Kapit-Mueller Hamiltonians have the same zero-energy ground state. 
Nevertheless, we will show that these two Hamiltonians are (slightly) different, and therefore have (slightly) different excited states.
We know that the continuum Hamiltonian is given by
\begin{equation}
\bar{H}_c = \sum_{\bq \in \text{BZ}}  F_{\text{LLL}} ({\bf q})  \ \bar\rho_{{\bf q}} \bar\rho_{-{\bf q}}
\label{SameWay}
\end{equation}
with
\begin{equation}
\begin{aligned}
F_{\text{LLL}} ({\bf q})& = |d_{\text{LLL}}(\bq)|^2 \\
& = \sum_{{\bf H}}  e^{- \frac12 ({\bf q}+{\bf H})^2 l^2} \\
&= e^{-\frac12 {\bf q}^2 l^2} \theta_3(i q_x; i N_y)  \theta_3(i q_y; i N_y)
\end{aligned}
\end{equation}
where $\theta_3(z;\tau)$ is the theta function of the third kind, defined as $ \theta_3(z;\tau) = \sum_{n} \exp(\pi i n^2 \tau + 2 \pi i n z)$. 

Now, since the results of Sec. II are valid for any choice of hoppings (as long as their phases are consistent with a uniform flux per plaquette of $n_{\phi}=P/Q$), they are in particular valid for the KM model. 
Therefore, we know that the KM Hamiltonian can be written in the same way as in Eq.~(\ref{SameWay}), except with a different $F(\bq)$ function:
\begin{equation}
\bar{H}_s = \sum_{\bq \in \text{BZ}}  F_s ({\bf q})  \ \bar\rho_{{\bf q}} \bar\rho_{-{\bf q}}
\end{equation}
For an on-site interaction, we simply have $F_s ({\bf q}) = |d_s(\bq)|^2$, where $|d_s(\bq)|^2$ is the form factor corresponding to the KM single-particle wave functions. 
Since we know these wave functions from Eq.~(\ref{sampled}), we can use the definition of the distance given in Eq.~(\ref{distance}) to find 
\begin{equation}
\begin{aligned}
F_s ({\bf q}) &=  |d_s(\bq)|^2 \\
& = e^{-\frac12 {\bf q}^2 l^2}  \theta_3\left(-\frac12(q_y - i q_x); i \frac{N_y}{2}\right)^2  \\
\times &\theta_3\left(-\frac12(q_x - i q_y); i \frac{N_y}{2}\right)^2.
\end{aligned}
\end{equation}
Numerically, we find that the two form factor $|d_{\text{LLL}}(\bq)|^2$ and $|d_s(\bq)|^2$, and therefore the two Hamiltonians $\bar{H}_c$ and $\bar{H}_s$, are very close to each other, but strictly different. 

The fact that two different interaction Hamiltonians can have the same Laughlin wavefunction as their ground-state manifold is peculiar to the torus and can easily be explained by looking at the two-particle problem. 
On the disk, inside each center-of-mass angular momentum sector, the two-particle Hamiltonian for the $m$th pseudopotential is given by a projector:
\begin{equation}
\mathcal{H}_{\text{Disk}} = \nu_m \ket{\Psi_m} \bra{\Psi_m} 
\end{equation}
where $ \ket{\Psi_m}$ is the only two-particle state with relative angular momentum $m$ inside the given center-of-mass sector.

On the torus, inside each center-of-mass $K_y^{\text{CM}}$ sector,  the two-particle Hamiltonian for a pseudopotential is given by a sum of two projectors\cite{PhysRevLett.111.126802}:
\begin{equation}
\mathcal{H} = \nu_{m,1} \ket{\Psi_{m,1}} \bra{\Psi_{m,1}} +  \nu_{m,2} \ket{\Psi_{m,2}} \bra{\Psi_{m,2}}
\end{equation}
with $ \nu_{m,1} \simeq \nu_{m,2}$.
If $\mathcal{U}$ is a unitary transformation that does not mix the space spanned by $ \{\ket{\Psi_{m,1}},  \ket{\Psi_{m,2}}\}$ and its complement, then $\mathcal{H}'=\mathcal{U}^{\dagger} \mathcal{H} \mathcal{U}$  will be another Hamiltonian with the same many-body zero-energy ground state.
Furthermore, the ground state obviously does not care about the values of $\nu_1$ and $\nu_2$ as long as they are strictly positive.
Since an overall multiplicative constant in front of the Hamiltonian can be disregarded, the relevant parameter is $\nu_2/\nu_1$.
We confirmed by a numerical calculation that $H_c$ and $H_s$ project onto the same two-dimensional Hilbert space, and we find that they have a different ratio $\nu_2/\nu_1$.
The two Hamiltonians $H_c$ and  $H_s$ therefore have the same zero-energy ground state but different excitations.
%

Naturally, since the results of Sec. II apply to the KM model, we know that its Berry curvature is exactly flat for our special system size. 
We should emphasize that this result is non trivial since the Berry curvature of this model is not flat for a generic system size.

%
%
%

\subsection{The impossibility of exactly flat Chern bands with strictly short-ranged hoppings}
The following theorem was recently shown\cite{1751-8121-47-15-152001}. Only up to two of the three following criteria can be met simultaneously: a model with (1) strictly short range hopping that exhibits a band that is (2) exactly flat and (3) that has a non zero Chern number. 
This is in apparent contradiction with our work that shows that the three criteria can be met simultaneously in the Hofstadter model for the right system size.
There is actually no contradiction since the theorem was shown for infinite-size systems, while our analysis only works for strictly finite-size systems.
Nevertheless, since we can reach arbitrarily large system sizes, our work constitutes somehow a way around this theorem.
We generalize the theorem as follows. Only up to three of the four following critieria can be met simultaneously: a model with (1) strictly short range hopping that exhibits a band that is (2) exactly flat and (3) that has a non zero Chern number and (4) that is defined on an infinite-size system.

%
%
%

%
\section{Generalization}


\subsection{The case of $|C|>1$}
The understanding of the physics of bands with Chern number one obtained in lattice models is based on the parallel with Landau levels since they have the same Chern number.
The existence of bands with $|C|>1$ in certain lattice models is therefore particularly promising as it could lead to new phases that are not present in Landau levels.
New FQH-resembling phases were numerically discovered in such bands at filling fractions $\nu=1/(|C|+1)$ for bosons and $\nu=1/(2|C|+1)$ for fermions \cite{PhysRevLett.103.105303,PhysRevLett.108.256809,PhysRevB.87.205137,PhysRevLett.110.106802, PhysRevB.89.155113,PhysRevLett.111.186804}.

As a reminder, the Chern number $C$ of each band is given by $PC \mod Q = 1$, where $n_{\phi}=P/Q$ with $P$ and $Q$ coprime. 
Let us now discuss the case of $P>1$, in which case $|C|$ can be larger than 1. 
The GMP operators $\bar\rho_{\bf q}$ depend on only one parameter $C$ that we will now write explicitly:  $\bar\rho_{\bf q}^{(C)}$. 
While for the case $C=1$ the reference systems are Landau levels, it is not obvious what the simplest continuum setting with a Chern $|C| > 1$ band is. 
It was argued in Refs. \cite{PhysRevLett.110.106802, PhysRevB.89.155113} that this system is a $C$-color quantum Hall system with color-entangled periodic boundary conditions. 
Each band of this system is made of a $N_x \times N_y$ BZ of color-entangled basis states $\ket{K_x,K_y}$. 
In the continuum, it is always possible to choose $N_x=1$, in which case the one-body basis is a set of $N_y$ states indexed by $K_y$.
In the present language, the Hamiltonian used in Refs.~\cite{PhysRevLett.110.106802, PhysRevB.89.155113} is given by:
\begin{equation}
\bar{H}_c  = \sum_{\bq \in \text{BZ}}  F_{\text{LLL}}^{(C)} ({\bf q})  \ \bar\rho_{{\bf q}}^{(C)}  \bar\rho_{-{\bf q}}^{(C)}, 
\label{PHPc}
\end{equation}
where
\begin{equation}
\begin{aligned}
F^{(C)}_{\text{LLL}}({\bf q}) &= |d^{(C)}_{\text{LLL}}({\bf q})|^2 \\
&= \sum_{{\bf H}} e^{- \frac12 C ({\bf q}+{\bf H})^2 l^2} 
\end{aligned}
\end{equation}
for the 0-th pseudopotential. 
The difference between $C=1$ and $|C|>1$ is therefore twofold: (1) the projected densities obey a $C$-GMP algebra and (2) the form factor is transformed according to $l^2 \rightarrow C l^2$.

The zero-energy ground states of $\bar{H}_c$ numerically obtained at filling fraction $\nu=1/(|C|+1)$ for bosons and $\nu=1/(2|C|+1)$ for fermions were baptized color-entangled Halperin states and were shown to be distinct from the SU($C$)-singlet Halperin state on the torus \cite{PhysRevLett.110.106802, PhysRevB.89.155113}.


As explained in Sec. II, the projected interaction Hamiltonian in the lattice case is given by
\begin{equation}
\bar{H}_{\text{Lat}} = \sum_{\bq \in \text{BZ}}  F_{\text{Lat}}^{(C)}({\bf q})  \ \bar\rho_{{\bf q}}^{(C)}  \bar\rho_{-{\bf q}}^{(C)} 
\label{PHP}
\end{equation}
with $F_{\text{Lat}}^{(C)}({\bf q}) = V({\bf q}) |d_0^{(C)}({\bf q})|^2$.

\begin{figure}
  \centering
  \begin{tabular}{@{}p{0.45\linewidth}@{\quad}p{0.45\linewidth}@{}}
    \subfigimg[width=\linewidth]{a)}{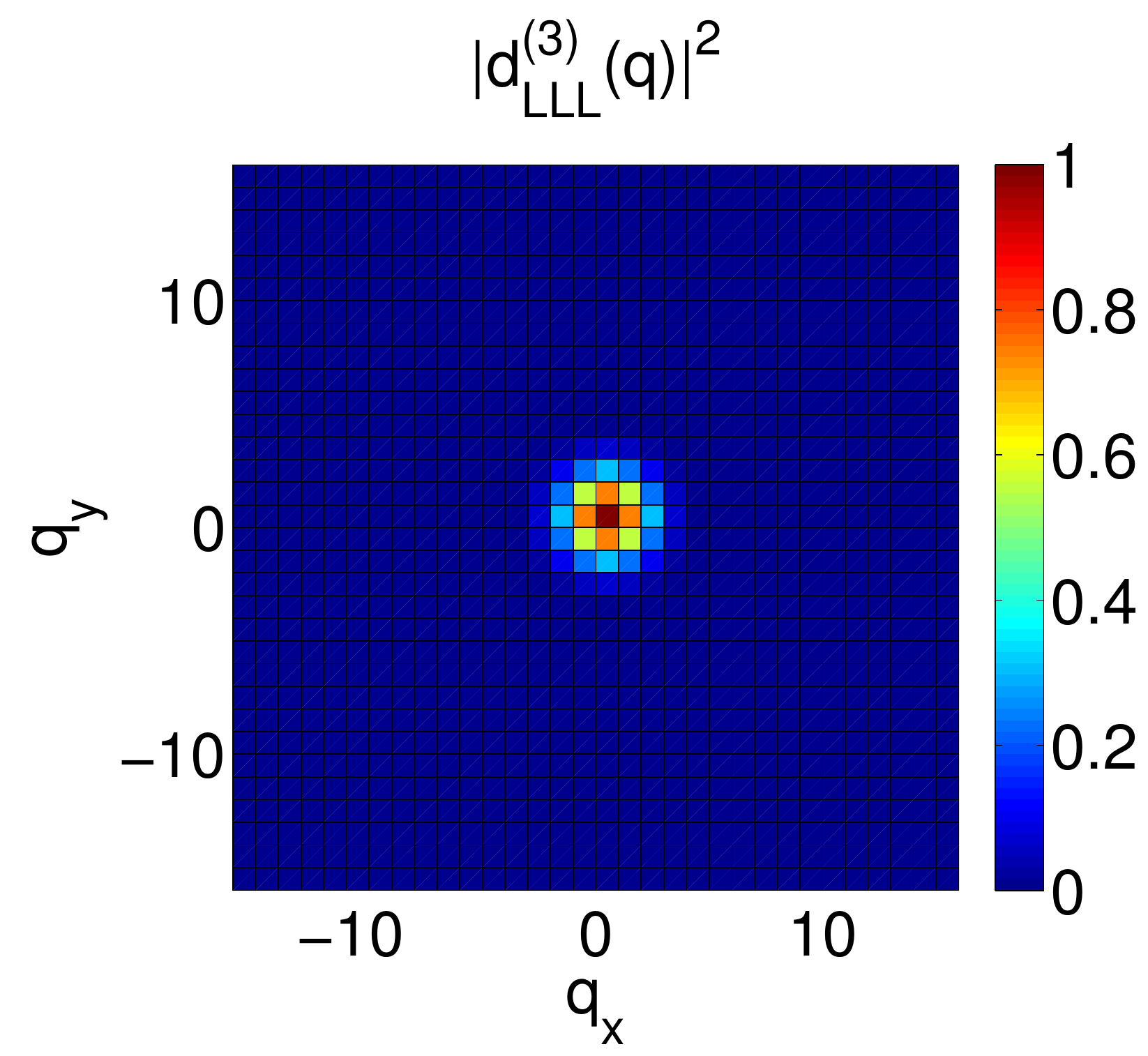} &
    \subfigimg[width=\linewidth]{b)}{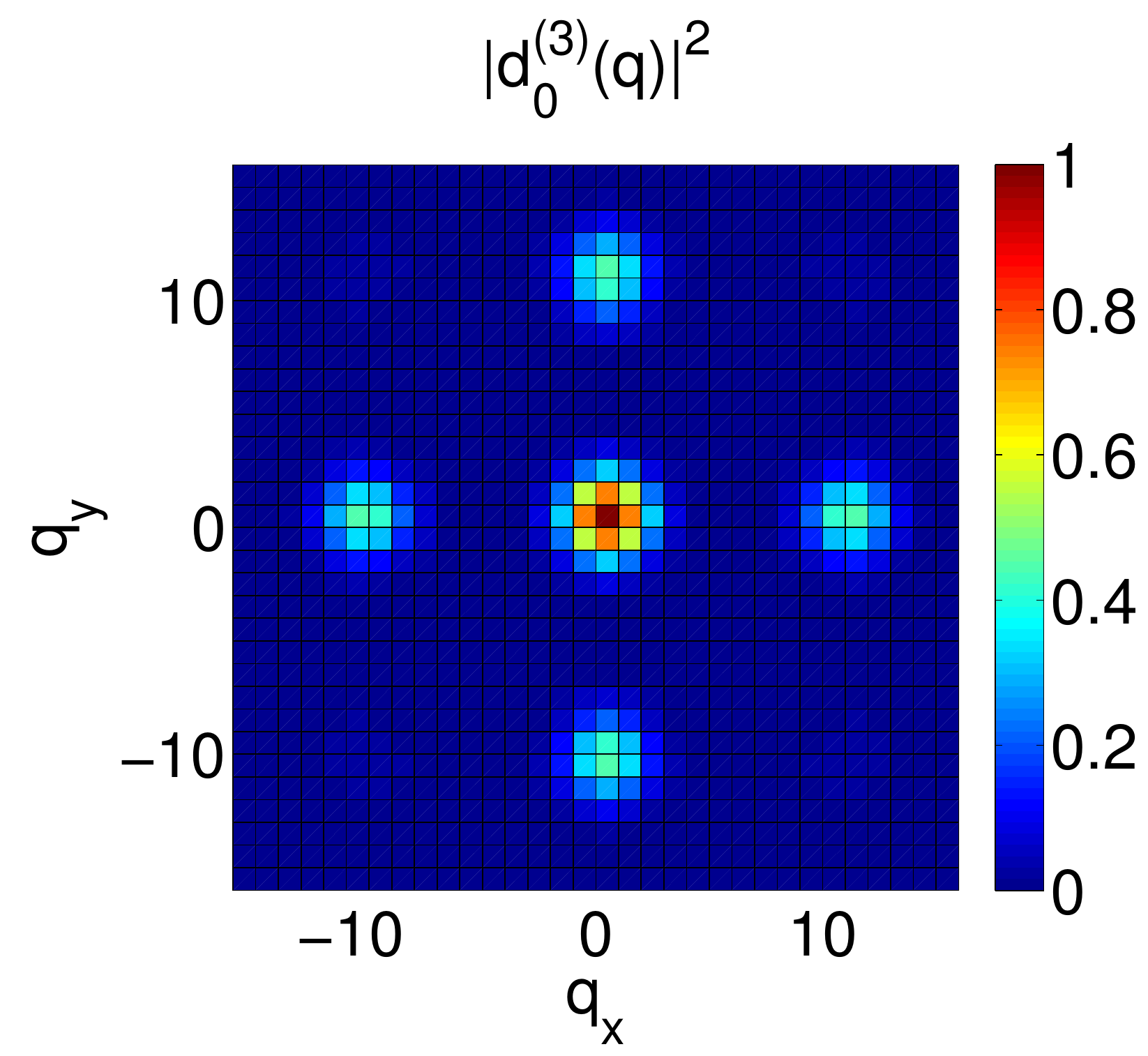} \\
  \end{tabular}
\caption{\label{FqC}
Form factors for $C=3$ in the LLL and lattice cases. 
The form factor in the lattice case is given for nearest-neighbor hopping. 
The flux density is $n_{\phi} = 11/32$ and $N_y=32$. 
The peaks centered at $(\pm N_y/3, \pm N_y/3)$ are too small to be seen clearly in (b).
}
\end{figure}

We find again that the only difference between the Hofstadter model and the continuum system lies in the difference between two functions: $|d_0^{(C)}({\bf q})|^2$ and $|d^{(C)}_{\text{LLL}}({\bf q})|^2$. 
The peculiar thing about the case $|C|>1$ is that the two functions are fundamentally different and this difference does not vanish in the large $Q$ limit.
As can be seen in Fig.~\ref{FqC}, $|d^{(C)}_{\text{LLL}}({\bf q})|^2$ is a periodicized Gaussian centered at $(0,0)$, while $|d_0^{(C)}({\bf q})|^2$ has, besides the same Gaussian centered at $(0,0)$, smaller Gaussian peaks centered at $(q_x,q_y)=(m_x N_y/C, m_y N_y/C)$ with $m_x, m_y$ integers. 
The difference between the Gaussians centered at $(0,0)$ for $|d^{(C)}_{\text{LLL}}({\bf q})|^2$ and $|d_0^{(C)}({\bf q})|^2$ goes like $1/Q$, but the satellite peaks of  $|d_0^{(C)}({\bf q})|^2$ do not vanish in the large $Q$ limit.

%
These peaks correspond to umklapp terms already discussed in the $C=2$ case in Ref~\cite{PhysRevLett.108.256809,PhysRevB.90.075104}.
They are caused by the rapid oscillation with period $C$ of the solutions of the Harper equation and are therefore peculiar to the Hofstadter model. In the case of $C=2$, it is possible to make these peaks approximately vanish by taking a nearest-neighbor interaction instead of an on-site interaction. The interaction potential then becomes
\begin{equation}
V({\bf q}) = 2 \left( \cos(\frac{q_x}{N_y} 2 \pi) + \cos(\frac{q_y}{N_y} 2 \pi) \right),
\end{equation}
which makes $F_{\text{Lat}}^{(C)}({\bf q})$ vanish at $(0, \pm \pi)$ and $(\pm \pi,0)$, i.e. the center of the umklapp peaks. 

The impact of these umklapp peaks for general $C$ is an open question that we will address in a future work.


\subsection{Higher bands}
We generalize the results of Sec. II to higher bands.
For the 0-th pseudopotential projected to the $n$-th continuum LL, we have
\begin{equation}
\bar{H}_c \equiv P_{\text{LL}}^{(n)} H_{\text{LL}} P_{\text{LL}}^{(n)} = \sum_{\bq \in \text{BZ}}  F_{\text{LL}}^{(n)} ({\bf q})  \ \bar\rho_{{\bf q}} \bar\rho_{-{\bf q}}
\end{equation}
with
\begin{equation}
\begin{aligned}
F^{(n)}_{\text{LL}}({\bf q}) &= |d^{(n)}_{\text{LL}}({\bf q})|^2 \\
&= \sum_{{\bf H}}  \left(L^n\left(\frac12 ({\bf q}+{\bf H})^2 l^2\right)\right)^2 e^{- \frac12 ({\bf q}+{\bf H})^2 l^2}
\end{aligned}
\end{equation}
where $L^n$ is the $n$-th Laguerre polynomial.

The interaction Hamiltonian projected to the $n$-th band of the Hofstadter problem is given by
\begin{equation}
\bar{H}_{\text{Lat}} \equiv P_{\text{Lat}}^{(n)} H_{\text{Lat}} P_{\text{Lat}}^{(n)} = \sum_{\bq \in \text{BZ}}  F_{\text{Lat}}^{(n)} ({\bf q})  \ \bar\rho_{{\bf q}} \bar\rho_{-{\bf q}}
\end{equation}
where $F_{\text{Lat}}^{(n)} ({\bf q}) = |d_n(\bq)|^2$ for an on-site interaction.
Since there is no closed analytical formula for $|d_n(\bq)|^2$, it was computed numerically. 
We find that the difference between $|d_n(\bq)|^2$ and$|d^{(n)}_{\text{LL}}({\bf q})|^2$ is small and decays like $1/Q$:
\begin{equation}
 |d_n(\bq)|^2 - |d^{(n)}_{\text{LL}}({\bf q})|^2 \propto \frac1{Q} \ll 1
\label{HigherBands}
\end{equation}
for $n$ sufficiently smaller than $Q/2$. 
This last condition comes from the fact that the Hofstadter model has Q bands but the bands $n$ and $Q-n$ are related by a symmetry given by (in the case of even $Q$)
\begin{equation}
u_{K_y,Q-n}(x) = u_{K_y,n}(x+Q/2) (-1)^x.
\end{equation}
The identification of the Hofstatder band index with the LL index can therefore only work for $n<Q/2$. 
Numerically, we find that the difference between the two form factors increases with $n$ and that the latter identification ceases to be valid for $n\gtrsim Q/5$.


 \subsection{Higher pseudo-potentials}
We now generalize the results of Sec. II to higher pseudopotentials.
The interaction Hamiltonian of the $m$th pseudopotential projected to the LLL is given by
\begin{equation}
\bar{H}_c \equiv P_{\text{LLL}} H_{\text{LLL}} P_{\text{LLL}} = \sum_{\bq \in \text{BZ}}  F_{\text{LLL}}^{(m)} ({\bf q})  \ \bar\rho_{{\bf q}} \bar\rho_{-{\bf q}},
\end{equation}
where 
\begin{equation}
F^{(m)}_{\text{LLL}}({\bf q}) = \sum_{{\bf H}}  L^m(({\bf q}+{\bf H})^2 l^2)  e^{- \frac12 ({\bf q}+{\bf H})^2 l^2}.
\end{equation}

Like in section II, we can find the lattice interaction that will lead to the desired Hamiltonian by doing the following:
\begin{equation}
V^{(m)}({\bf q}) = \frac{F^{(m)}_{\text{LLL}}({\bf q})}{|d_0({\bf q})|^2}\text{.}
\label{VqEq}
\end{equation}
The range of the interaction will typically be of the order of the magnetic length, which goes like $\sqrt{Q}$.

\begin{figure}
\begin{center}
\includegraphics[width=0.45\columnwidth]{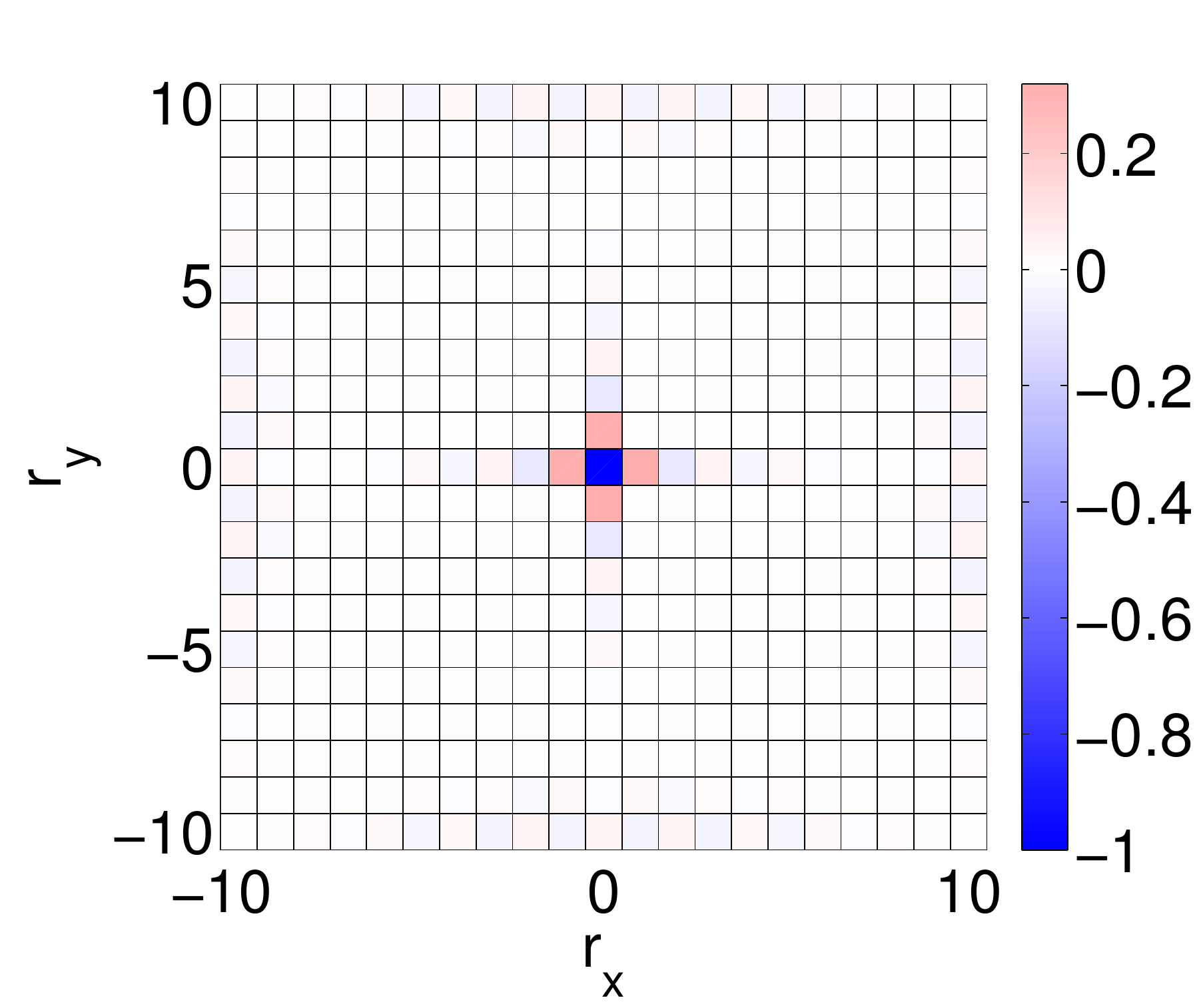}
\caption{\label{VrPP1}
Lattice interaction $V^{(1)}(\br)$ reproducing the $m=1$ pseudopotential. 
The form factor $|d_0(\bq)|^2$ was calculated for a nearest-neighbor hopping of 1 and a NNN hopping of 0.27. 
The plot is given for $N_y = 21$.
}
\end{center}
\end{figure}

We give a plot of $V^{(1)}({\bf r})$ in Fig.~\ref{VrPP1} in the case of $Q=21$.
The potential is minimal for $\br=0$ (it takes a value of $-1$), then takes its maximal value of $\simeq 0.3$ for nearest neighbor interaction and finally decreases over a few sites.
This behavior is easily explained by the following argument.
We can focus on the small ${\bf q}$ behavior, for which we can forget the sum over ${\bf H}$ and write
\begin{equation}
 V^{(m)}({\bf q})  \simeq   L^m(({\bf q})^2 l^2) \equiv 1 - {\bf q}^2 l^2
\end{equation}
since we know that $|d_0({\bf q})|^2 \simeq  e^{- \frac12 ({\bf q})^2 l^2} $ for small $\bq$.

We can approximate this potential with a simple lattice interaction by doing the following:
\begin{equation}
 1 - {\bf q}^2 l^2 \simeq 2 \left(\cos({\bf q}_x l) + \cos({\bf q}_y l) \right) - 3 
\end{equation}
where 
\begin{equation}
 {\bf q}_{x,y} l = \frac{q_{x,y}}{Q} 2 \pi \sqrt{\frac{Q}{2 \pi}} 
\end{equation}
and $q_{x,y}$ are integers.
If we round $\sqrt{\frac{Q}{2 \pi}}$ to the closest integer $r$, the interaction potential can be approximated by 
\begin{equation}
\begin{aligned}
 V(x, y) &= V_0 (-3 \delta_{x,0} \delta_{y,0} \\
&+ \delta_{x,-r} \delta_{y,0} + \delta_{x,r} \delta_{y,0}  +  \delta_{x,0} \delta_{y,r} + \delta_{x,0} \delta_{y,-r}). 
\end{aligned}
\end{equation}
This interaction reproduces fairly well the behavior found numerically in Fig~\ref{VrPP1}, since $\sqrt{Q/2\pi} \simeq 1.8$.

\subsection{Trapping potential}
In this section, we discuss the addition to the non interacting Hofstadter model $H_0$ of a periodicized version of a harmonic trap given by $H_W = \sum_{x,y}  H_W(x,y) \ket{x,y}\bra{x,y}$ with
\begin{equation}
\begin{aligned}
H_W(x,y) &=- 2 \left(W_x \cos\left(2 \pi \frac{x}{Q}\right) + W_y \cos\left(2 \pi \frac{y}{N_y}\right)\right)\text{.}
\end{aligned}
\end{equation}
We can write it in terms of unprojected density operators:
\begin{equation}
H_W = - W_x (\tilde\rho_x + \tilde\rho_{-x}) -W_y (\tilde\rho_y + \tilde\rho_{-y})\text{.}  
\end{equation}

If $W_x$ and $W_y$ are small enough compared to the band gap (i.e., compared to $t \equiv t_x = t_y$), we can use degenerate perturbation theory, which at first order in this case, amounts to diagonalizing $H_W$ projected to each band separately, $\bar{H}_W^{(n)} \equiv P_n H_W P_n$.
Starting from $H \ket{K_y, n} = E_n \ket{K_y, n}$ at zeroth order, we obtain
\begin{equation}
(H+H_W) \ket{l, n} \simeq (E_n+\epsilon_l^{(n)}) \ket{l, n}
\end{equation}
where $\bar{H}_W^{(n)} \ket{l, n} = \epsilon_l^{(n)} \ket{l, n}$ and $\epsilon_l^{(n)} \ll E_n$ with $l$ indexing the eigenvalues in ascending order.
The trapping potential will therefore give a small dispersion to each band and rearrange the states inside each band from $\{\ket{K_y}\}$ to $\{\ket{l}\}$.

Now, in order to diagonalize $\bar{H}_W^{(n)}$, we should find its matrix elements in the $\{\ket{K_y}\}$ basis.
Taking advantage of the projection of $\tilde{\rho}_{\bq}$ derived in Section II, we find
\begin{equation}
\bar{H}_W^{(n)} = -d_n(1,0) \left(  W_x (\bar\rho_x + \bar\rho_{-x}) +W_y (\bar\rho_y + \bar\rho_{-y}) \right)
\end{equation}
where we used $t_x=t_y$ to write $d_n(1,0) = d_n(0,1)$.
In the case $C=1$ and $W \equiv W_x = W_y$, this leads to an appealingly symmetric notation:
\begin{equation}
\begin{aligned}
H_0 &= -t (\hat{T}_x + \hat{T}_{-x} +\hat{T}_y + \hat{T}_{-y}) \\
\bar{H}_W^{(n)} &= - \tilde{W} (T_x + T_{-x}+ T_y + T_{-y}) 
\end{aligned}
\end{equation}
where $\tilde{W} = W d_n(1,0)$.

Interestingly, if we identify the set of $\ket{K_y}$ states with a putative position eigenstate $\ket{\tilde{x}=0, \tilde{y}=K_y}$ on a 1 by $N_y$ lattice, $\bar{H}_W^{(n)}$ is a Hofstadter model with a flux per plaquette of $\tilde{n}_{\phi}=C/N_y$ and with hoppings given by $\tilde{t}_x=d_n(1,0) W_x$ and $\tilde{t}_y=d_n(1,0) W_y$ since the following relations
\begin{equation}
\begin{aligned}
\bar\rho_y  \ket{K_y} &= \ket{K_y + 1} \\
\bar\rho_x  \ket{K_y} &=  e^{-i \frac{2 \pi}{N_y} C K_y}  \ket{K_y}
\end{aligned}
\end{equation}
give effective ``hoppings'' with the appropriate phase in this putative real space.

Since, for our system size, $N_x =1$, in terms of the putative real space all the states have $\tilde{x}=0$. 
The boundary conditions are periodic in the BZ and the ``hopping'' in the $x$ direction in momentum space $\bar\rho_x$ therefore takes a state to itself, only with a phase chosen according to the flux density $\tilde{n}_{\phi}$.
In other words, $\bar{H}_W^{(n)}$ is a Hofstadter model with only one vertical unit cell of $N_y$ plaquettes.


As $\bar{H}_W^{(n)}$ is a Hofstadter Hamiltonian with nearest-neighbor hopping, its spectrum will also be given by the Harper equation, just like the initial one-body Hamiltonian $H_0$.
Higher harmonics in the potential $W(x,y)$ would correspond to longer-range hopping in the putative real space.

In the case of $P=1$, we also have $C=1$, which means that $\tilde{n}_{\phi} = n_{\phi}$. In that case, assuming again $W_x = W_y$, the splitting of each band is easily found in terms of $E_n$, the spectrum of $H_0$:
\begin{equation}
\epsilon_l^{(n)} = \frac{\tilde{W}}{t} E_l.
\end{equation}

\begin{figure}
\begin{center}
\includegraphics[width=0.45\columnwidth]{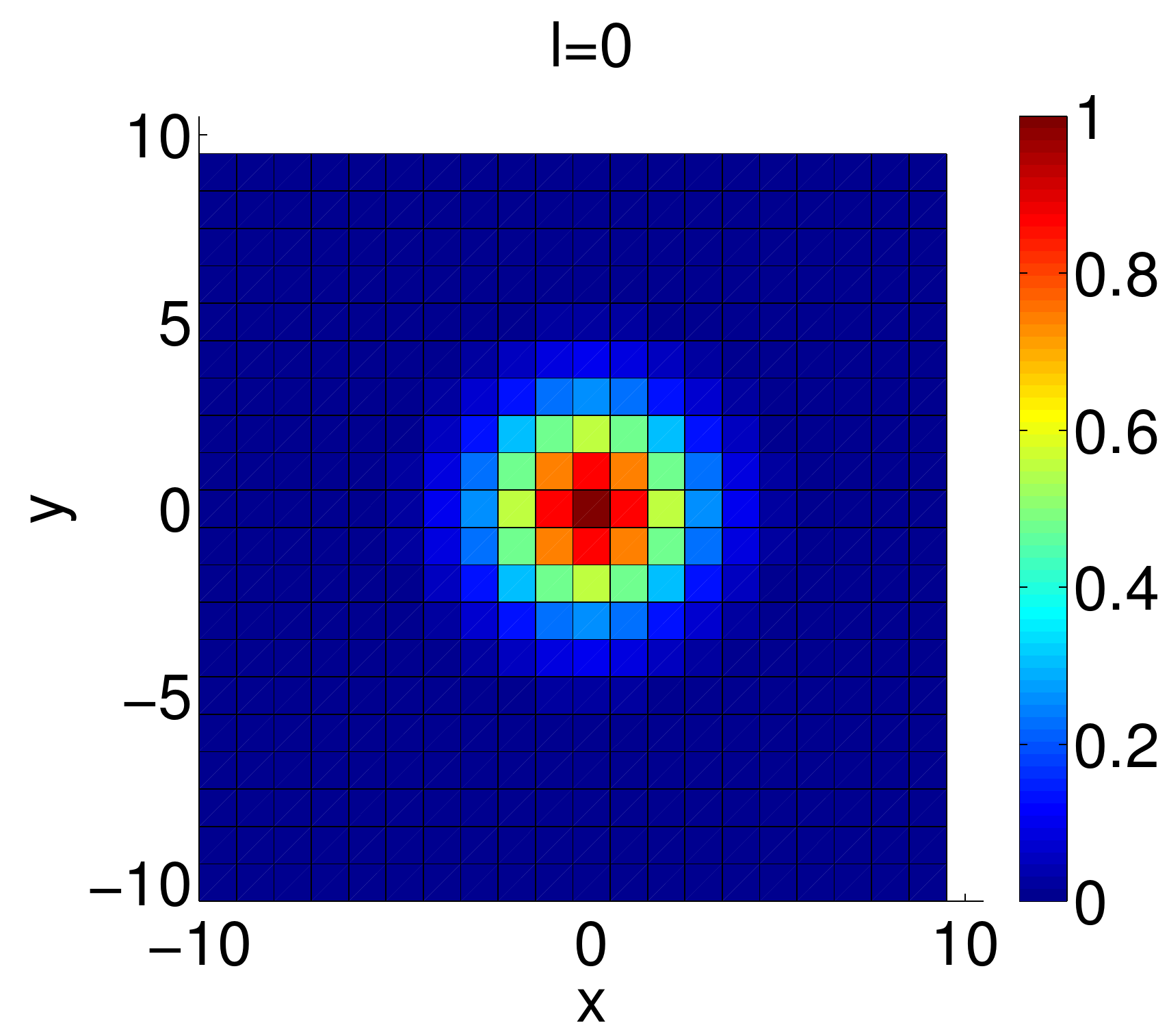}
\includegraphics[width=0.45\columnwidth]{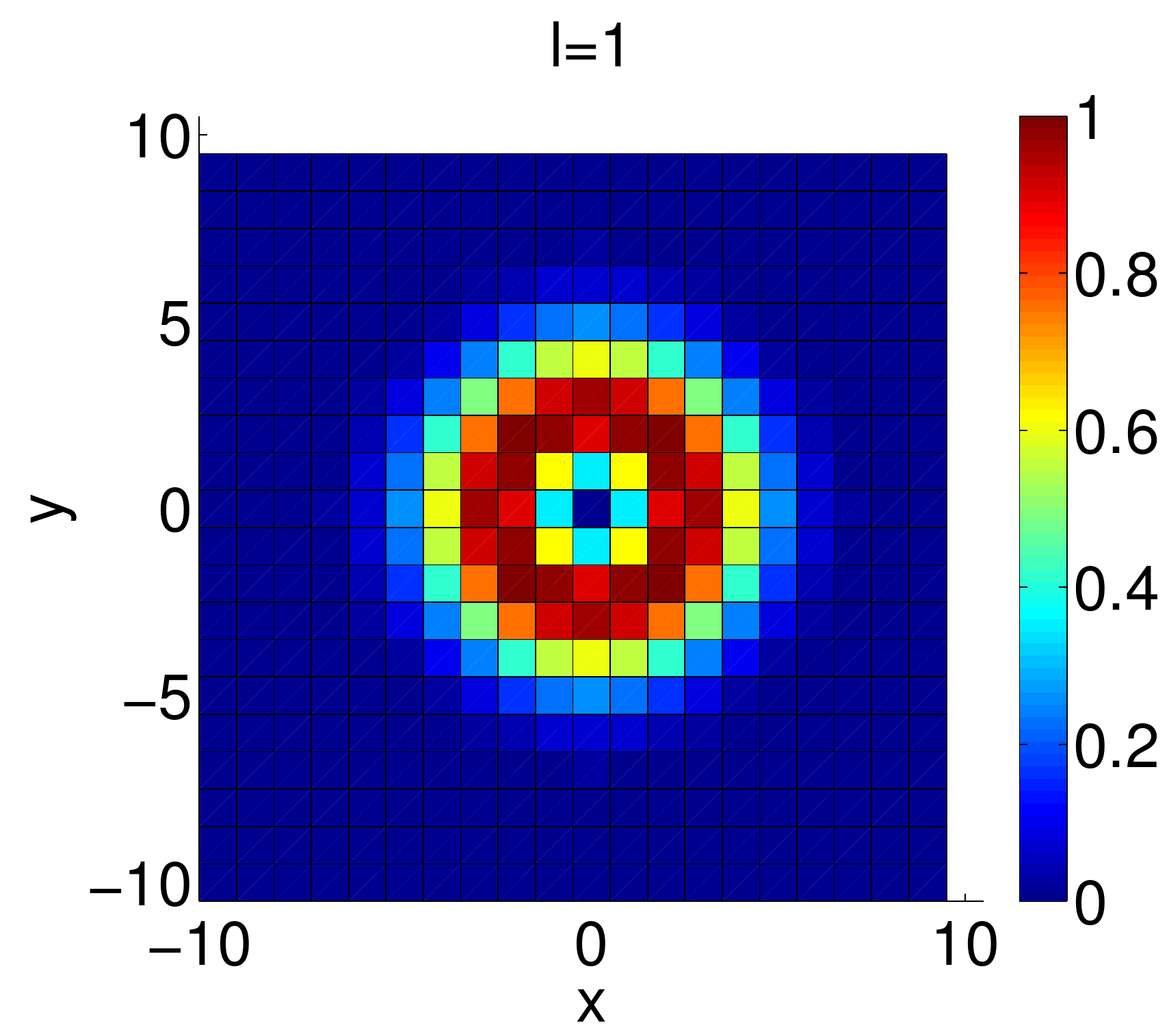}
\caption{\label{Leigs}
Real space wavefunction of the trapping potential Hamiltonian eigenstates: $|\psi_l(\br)|^2$ in the case of $C=1$ and $N_y=21$.
}
\end{center}
\end{figure}

Without surprise, we find numerically that the lattice real-space wave functions $\psi_l(\br)$ in the lowest band resemble the toroidal equivalent of the angular momentum eigenstates given by $\psi_L(z) \sim z^L e^{-|z|^2/2l_B^2}$ on the disk with $l=L$, as can be seen in Fig.~\ref{Leigs}.
Due to the PBCs, this identification only works for $l < Q/2$, since the state $\ket{Q-l}$ is related to the state $\ket{l}$ by a translation of the center by $(N_y/2, N_y/2)$.

\begin{figure}
  \centering
  \begin{tabular}{@{}p{0.45\linewidth}@{\quad}p{0.45\linewidth}@{}}
    \subfigimg[width=\linewidth]{a)}{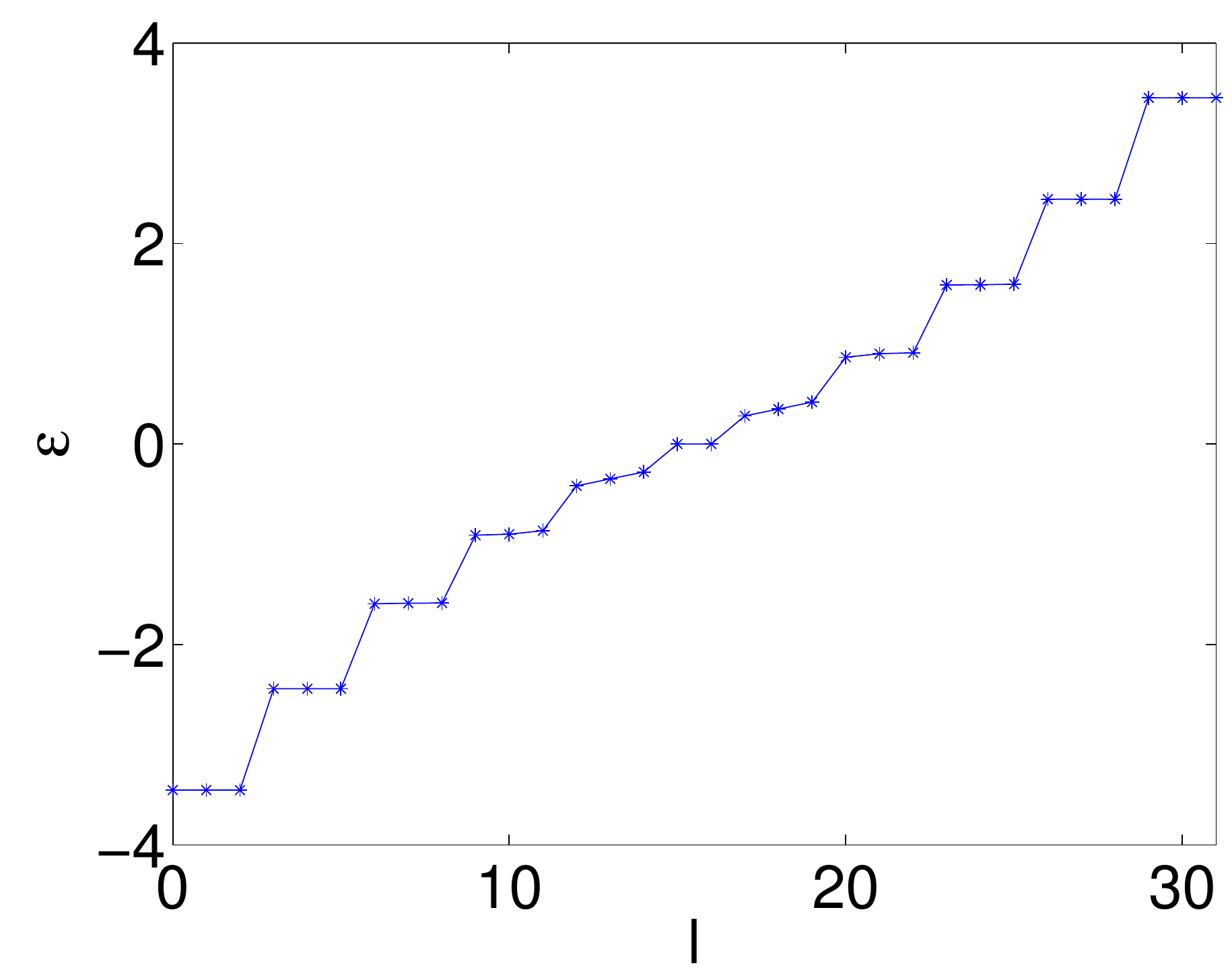} &
    \subfigimg[width=\linewidth]{b)}{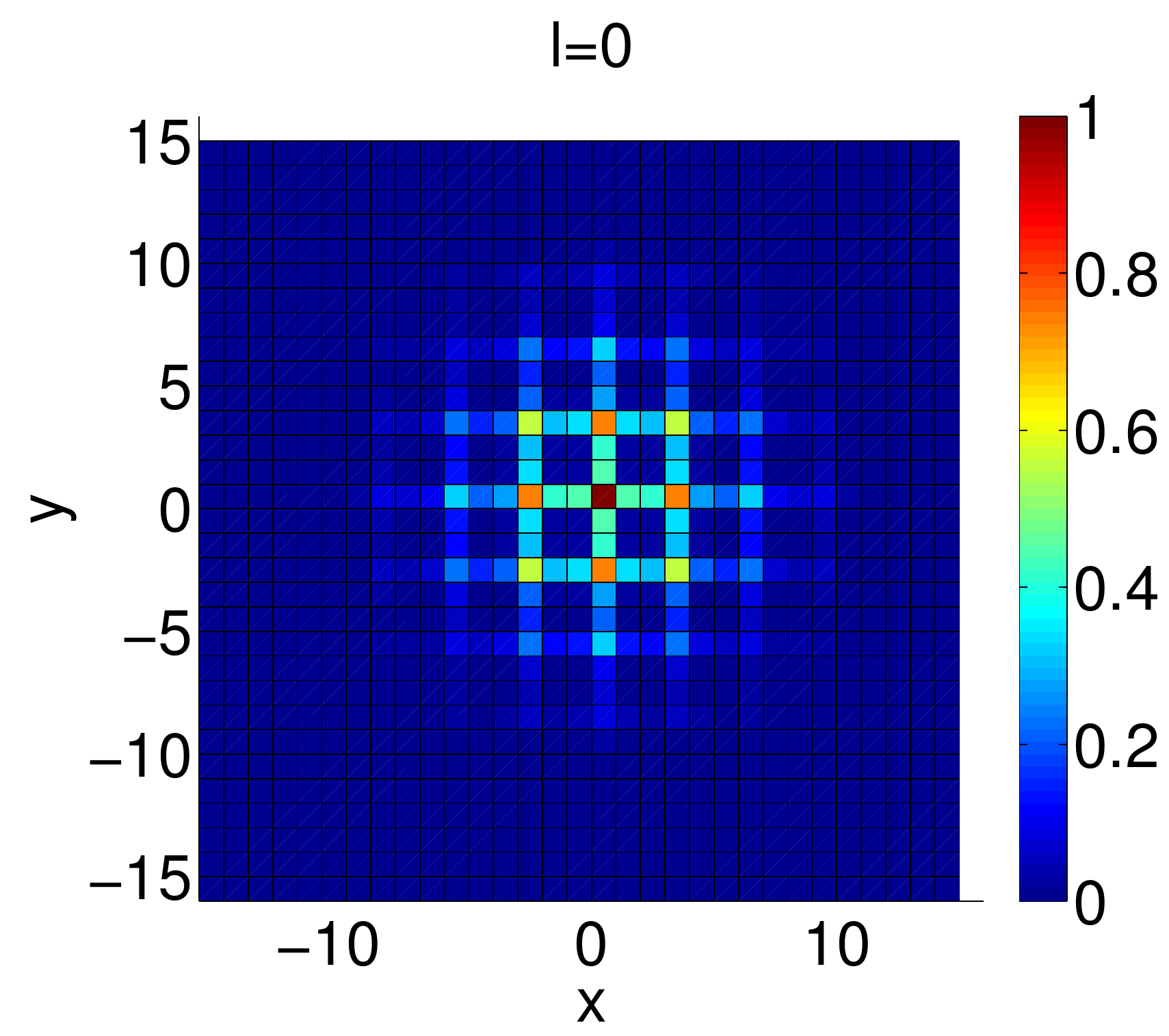} \\
    \subfigimg[width=\linewidth]{c)}{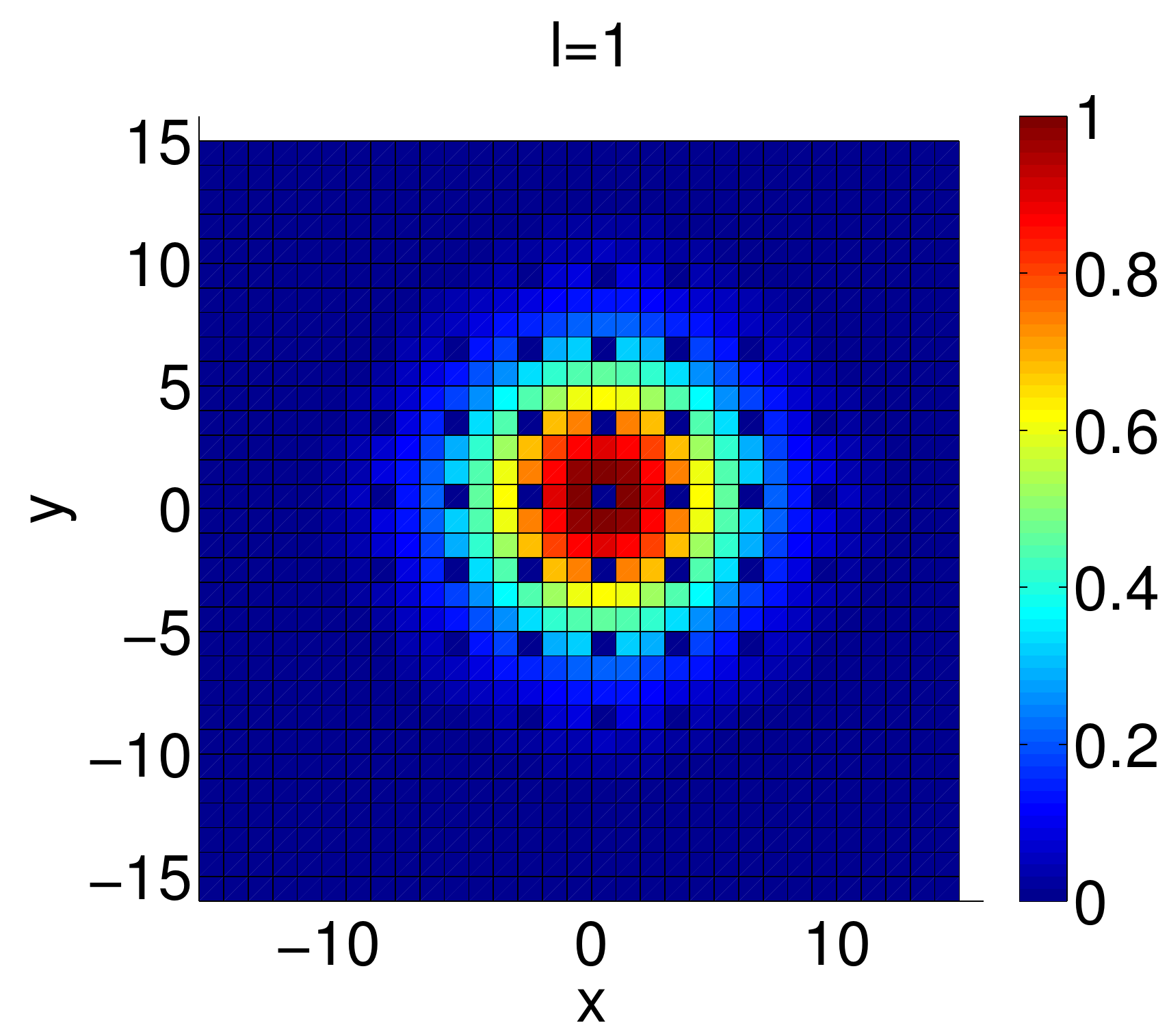} &
    \subfigimg[width=\linewidth]{d)}{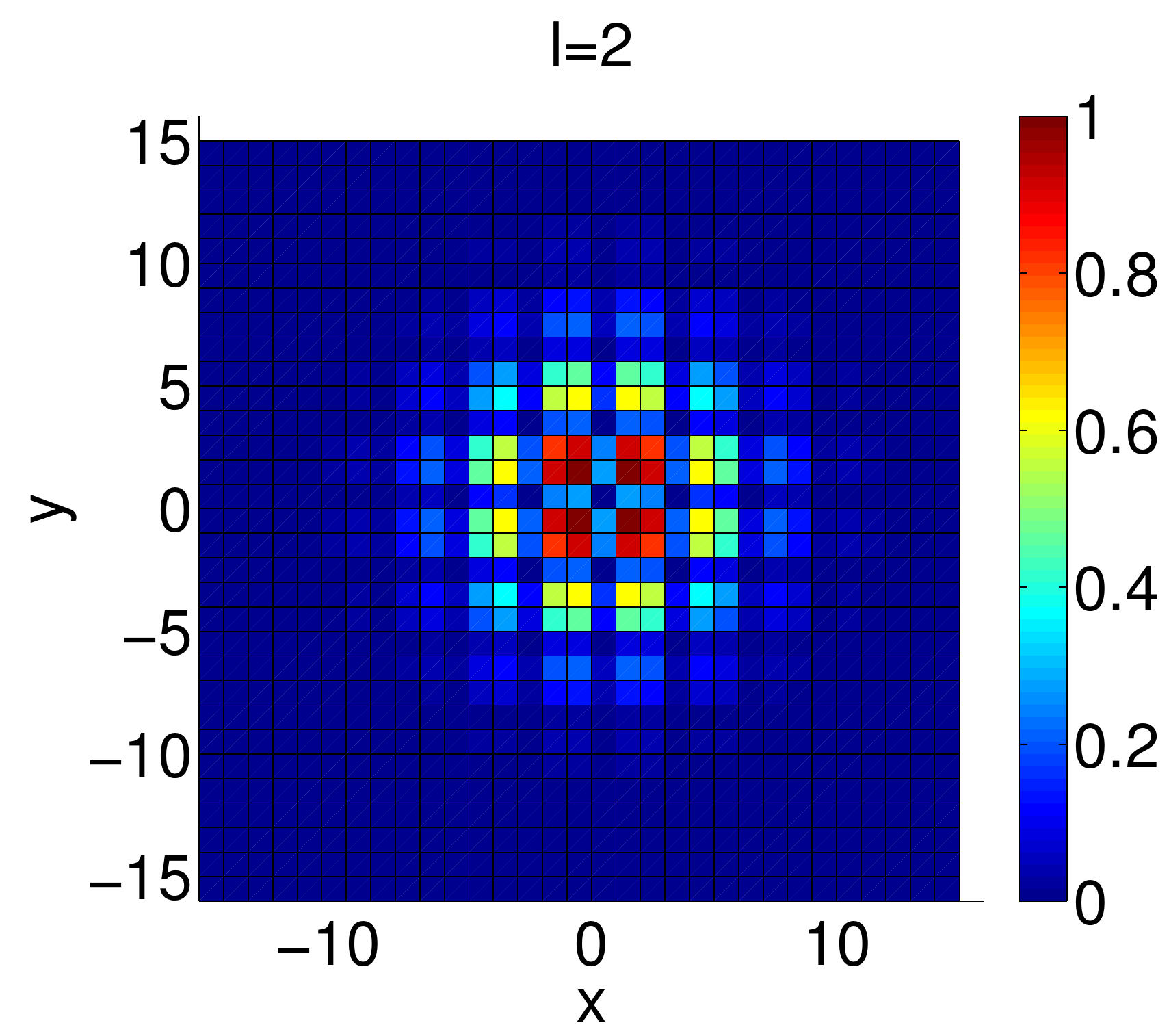}
  \end{tabular}
\caption{\label{CLSpectrum}
(a) Normalized spectrum $\epsilon_l^{(n)}/\tilde{W}$ of the trapping potential projected to one band with Chern number $C=3$.
(b)--(d) Real-space wavefunction $|\psi_l(\br)|^2$ of the first three eigenstates, corresponding to $L=0$ and $\sigma=0,1,2$. The system size is given by $N_y=32$ and the flux density is $n_{\phi}=11/32$.
}
\end{figure}

We now turn to the case of $P>1$, for which $|C|>1$ and $\tilde{n}_{\phi}=C/N_y$.
Typically, we will be interested in the case of $|C|$ fixed at some relatively small integer and $N_y$ taken large.
The spectrum of Harper's equation is well known in this limit: it will tend to equally spaced values as $N_y$ goes to infinity, like Landau levels in the continuum, except that each of these values will be $C$ times degenerate. 
For finite $N_y$, the values are not exactly equally spaced and the $C$ degeneracy is only approximate [see Fig.~\ref{CLSpectrum}(a)].
Inside one approximately degenerate subspace, the $C$ wavefunctions will correspond to the same value of ``angular momentum'' $L$ but they will differ by a $C$ periodicity on the lattice [see Figs.~\ref{CLSpectrum}(b)--(d)]. 
The Hilbert space can then be reinterepreted as $\simeq N_y / C$ angular momentum values times $C$ different species.
The identification with the disk angular momentum eigenstates therefore becomes $l = CL + \sigma$ where $\sigma=0,\dots,C-1$ is the species index.
We recover the well-known result that the dynamics inside a Chern $C$ band can be interpreted in terms of $C$ different species\cite{PhysRevLett.103.105303, PhysRevLett.108.256809, PhysRevLett.110.106802, PhysRevB.89.155113}.


The symbol $\simeq$ was used since $N_y$ is generally not divisible by $C$. 
As seen in Fig.~\ref{CLSpectrum}(a), the $C$ degeneracy becomes worse as $l$ approaches $Q/2$ and the fact that $N_y$ is not divisible $C$ is accommodated by a few states near $Q/2$ for which the $C$ degeneracy is reduced.

Finally, the fact that a potential in real space gives, once projected to a given band, a system equivalent to a magnetic field in momentum space has been reported in previous work\cite{2014arXiv1403.6041P}. 
In this picture, the Berry curvature corresponds to the magnetic field in momentum space. 
A flat Berry curvature therefore corresponds to a constant flux density, as we find in our case.
Furthermore, the total flux of this putative magnetic field is obviously given by the Chern number since we have $N_y$ plaquettes with a flux per plaquette of $\tilde{n}_{\phi} = C/N_y$.
\section{Conclusion}
In conclusion, we have shown that all the bands of the Hofstadter model with flux density $P/Q$ have an exactly flat dispersion and Berry curvature when a system size of $Q$ by $Q$ plaquettes and periodic boundary conditions are chosen.
This constitutes a simple rule for choosing a particularly advantageous system size when designing a Hofstadter system whose size is controllable, like a qubit lattice\cite{PhysRevA.87.062336} or an optical cavity array\cite{PhysRevA.84.043804}.



We could then calculate the projection of the density operators to any of the bands.
The projected density operators were shown to obey exactly the same algebra as in the continuum, the so-called GMP algebra.
In a future work, techniques to estimate the many-body gap based on this algebra could be transposed to the Hofstadter model: the single-mode approximation\cite{PhysRevB.33.2481} and the Hamiltonian theory of composite fermions \cite{RevModPhys.75.1101, PhysRevB.86.195146}.

Furthermore, we calculated the lattice interaction Hamiltonian projected to any of the flat bands and showed explicitly its relation to pseudopotential Hamiltonians projected to a Landau level.
We isolated the difference between these two Hamiltonians to a very simple quantity: the difference between two real scalar form factors defined on a $Q$ by $Q$ BZ and giving the ``distances'' inside the flat band manifold of states.
The only difference between the lattice and the continuum system is therefore contained in these ``distances'' that are shown to be very similar but not strictly equal.
We then gave the lattice interaction that makes these two Hamiltonians strictly equal, therefore providing an exact zero-energy ground state for the Hofstadter model with the latter interaction.

This work was next discussed at the light of similar previous results: the Wannier construction for FCIs\cite{PhysRevLett.107.126803}, the Kapit-Mueller model\cite{PhysRevLett.105.215303}, and a theorem stating the impossibility for a strictly short range hopping model to host Chern bands with an exactly flat dispersion and Berry curvature \cite{1751-8121-47-15-152001}.

Finally, we generalized these results in several ways. 
First, we treated the case of higher Chern number, for which we showed that the difference between the Hofstadter model and its continuum counterpart, embodied in satellite peaks in the form factor corresponding to umklapp terms, does not vanish in the large $Q$ limit.
Second, we provided a straightforward generalization of our result to the case of higher bands and pseudopotentials. 
We showed that the first few bands of the Hofstadter model can be identified with the first few Landau levels and that higher pseudopotentials can be reached by using longer range lattice interactions.
Third, we discussed the addition of a harmonic trapping potential and exhibited an appealing symmetry of this problem in terms of a Hofstadter model in momentum space.


%

%
%

%

\begin{acknowledgements}
Helpful conversations with Gunnar M{\"o}ller, Alexander Seidel, and Fenner Harper are acknowledged. This work is supported by EPSRC Grant Nos. EP/I032487/1 and EP/I031014/1, the Clarendon Fund Scholarship, the Merton College Domus and Prize Scholarships, and the University of Oxford.
\end{acknowledgements}

\appendix

\section{Continuum wave functions sampled on the lattice}
In this Appendix, we derive the expression for the sampled Landau level single-particle wave functions with periodic boundary conditions.
Starting from the continuum wavefunction given in Eq.~(\ref{cont}),
\begin{equation} 
\begin{aligned}
&\psi^c_{\mathbf{K}}(\tilde{x}, \tilde{y}) = \sum_{n \in \mathbb{Z}} \sum_{m=0}^{N_x-1} \exp\left[ i \frac{2 \pi}{L_y} (K_y+ m N_y + n N_{\phi}) \tilde{y}  \right] \times \\
& \exp\left[i 2 \pi m \frac{K_x}{N_x}\right] \exp\left[ -\frac{\left(\tilde{x} + l^2 (K_y + m N_y + n N_{\phi}) \frac{2 \pi}{L_y}\right)^2  }{2 l^2} \right] \text{,}
\end{aligned}
\end{equation}
we substitute $\tilde{x}=x L_x/N_x Q$ and $\tilde{y}=y L_y/N_y$ with $x,y \in \mathbb{Z}$ to obtain
\begin{equation} 
\begin{aligned}
\psi^c_{\mathbf{K}}(x, y) &= \sum_{n \in \mathbb{Z}} \sum_{m=0}^{N_x-1}\exp\left[i 2 \pi m \frac{K_x}{N_x}\right] \times \\
& \exp\left[ i \frac{2 \pi}{L_y} (K_y+ m N_y + n N_{\phi}) \frac{y L_y}{N_y}  \right] \times \\
&  \exp\left[ -\frac{\left(\frac{x L_x}{N_x Q} + l^2 (K_y + m N_y + n N_{\phi}) \frac{2 \pi}{L_y}\right)^2  }{2 l^2} \right]. 
\end{aligned}
\end{equation}

In the argument of the second exponential, the terms $m N_y$ and $n N_{\phi}$ can be discarded since they are mutiples of $i 2 \pi$. 
Replacing $l^2$ by $L_x L_y / 2 \pi N_{\phi}$, the argument of the third exponential becomes
\be
\left[ - \frac12 2\pi \frac{L_x}{L_y} \frac{N_y}{N_x} \frac1{Q^2}\left(x+ \frac{Q}{N_{y}} (K_y + m N_y + n N_{\phi}) \right)^2  \right] \text{.}
\ee

Now, if we choose the plaquettes to be squares of side $a$, we have
\be
\frac{L_x}{L_y} \frac{N_y}{N_x} = \frac{N_x Q a}{N_y a} \frac{N_y}{N_x} = Q.
\ee

We finally obtain
\begin{equation} 
\begin{aligned}
&\psi^s_{\mathbf{K}}(x, y) =\sum_{n \in \mathbb{Z}} \sum_{m=0}^{N_x-1} \exp\left[ i \frac{2 \pi}{N_y} K_y y  \right] \exp\left[i 2 \pi m \frac{K_x}{N_x}\right] \times\\
& \exp\left[ -\frac12 2 \pi \frac{1}{Q} \left(x + \frac1{N_y} Q (K_y + m N_y + n N_{\phi}) \right)^2   \right] 
\end{aligned}
\end{equation}
which, by doing the change of variable $(m+n N_x) \rightarrow m$, is shown to be equal to the expression for the sampled wavefunctions given in the main text (Eq.~\ref{sampled}):
\begin{equation} 
\begin{aligned}
&\psi^s_{\mathbf{K}}(x, y) = \sum_{m \in \mathbb{Z}} \exp\left[ i \frac{2 \pi}{N_y} K_y y  \right] \exp\left[i 2 \pi m \frac{K_x}{N_x}\right] \times \\
& \exp\left[ -\frac12 2 \pi \frac{1}{Q} \left(x + \frac1{N_y} Q (K_y + m N_y ) \right)^2   \right] \text{.}
\end{aligned}
\end{equation}

\section{Overlaps of the sampled wavefunctions}
In this Appendix, we calculate the overlaps of the sampled wavefunctions given in Eq.~\ref{sampled} and derived in Appendix A.
These overlaps are written as $\sand{\psi^s_{\mathbf{K}}}{\psi^s_{\mathbf{K}'}} = \sum_z (\psi^{s}_{\mathbf{K}}(z))^* \psi^{s}_{\mathbf{K}'}(z)$ where $\psi^s_{\mathbf{K}}$ are the sampled continuum wavefunctions, defined for $K_{x,y} = 0, \dots, N_{x,y}$ and given by 
\begin{equation} 
\begin{aligned}
&\psi^s_{\mathbf{K}}(x, y) = \sum_{m \in \mathbb{Z}} \exp\left[ i \frac{2 \pi}{N_y} K_y y  \right] \exp\left[i 2 \pi m \frac{K_x}{N_x}\right] \\
& \exp\left[ -\frac12 2 \pi \frac{1}{Q} \left(x + \frac1{N_y} Q (K_y + m N_y ) \right)^2   \right] \text{.}
\end{aligned}
\end{equation}
These wavefunctions are Bloch waves:
\begin{equation}
\begin{aligned}
\psi^s_{\mathbf{K}}(x, y+1) = \psi^s_{\mathbf{K}}(x, y) e^{i \frac{2 \pi}{N_y} K_y}\\
\psi^s_{\mathbf{K}}(x+Q, y) = \psi^s_{\mathbf{K}}(x, y) e^{i 2 \pi \frac{K_x}{N_x}}.
\end{aligned}
\end{equation}
It is therefore clear that $\sand{\psi^s_{\mathbf{K}}}{\psi^s_{\mathbf{K}'}} \propto \delta_{K_{y}, K'_{y}}  \delta_{K_{x}, K'_{x}} $.

Let us now calculate the norm of these states:
\begin{equation}
\begin{aligned}
&\mathcal{N}^2(\mathbf{K}) = \sand{\psi^s_{\mathbf{K}}}{\psi^s_{\mathbf{K}}} = \sum_x  \sum_{m \in \mathbb{Z}} \sum_{m' \in \mathbb{Z}} \\
& \exp\left[-i 2 \pi m \frac{K_x}{N_x}\right]  \exp\left[ -\frac12 2 \pi \frac{1}{Q} \left(x + \frac1{N_y} Q (K_y + m N_y ) \right)^2   \right] \times \\
&   \exp\left[i 2 \pi m' \frac{K_x}{N_x}\right]  \exp\left[ -\frac12 2 \pi \frac{1}{Q} \left(x + \frac1{N_y} Q (K_y + m' N_y ) \right)^2   \right]. 
\end{aligned}
\end{equation}
From this expression, we can deduce that, in general, $\mathcal{N}^2(K_x+1, K_y) \neq \mathcal{N}^2(K_x, K_y)$ and $\mathcal{N}^2(K_x, K_y+1) \neq \mathcal{N}^2(K_x, K_y)$. 
Nevertheless, in the particular case of $N_x=1$, $N_y = Q$, this expression becomes
\begin{equation}
\begin{aligned}
\mathcal{N}^2(\mathbf{K}) &= \sand{\psi^s_{\mathbf{K}}}{\psi^s_{\mathbf{K}}} = \sum_x  \sum_{m \in \mathbb{Z}} \sum_{m' \in \mathbb{Z}}\\
&    \exp\left[ -\frac12 2 \pi \frac{1}{Q} \left(x + (K_y + m N_y ) \right)^2   \right] \times\\
&   \exp\left[ -\frac12 2 \pi \frac{1}{Q} \left(x +  (K_y + m' N_y ) \right)^2   \right] 
\end{aligned}
\end{equation}
which clearly does not depend on $K_y$ since one can always translate the sum over $x$. Since for $N_x=1$, all the states have the same $K_x$, this means that, in this particular case, all the states have the same norm.




\
\bibliography{hofstadter}

\end{document}